\documentclass{aa}  

\usepackage{graphicx}
\usepackage{txfonts}
\usepackage{amsmath,amsfonts,amssymb}
\usepackage{graphicx}
\usepackage{setspace}
\usepackage{tocloft}
\usepackage{subfig}
\usepackage[colorlinks, citecolor=blue, linkcolor=blue]{hyperref}
\usepackage[table]{xcolor}
\usepackage[normalem]{ulem}

\begin{document}

   \title{The CARMENES search for exoplanets around M dwarfs}

   \subtitle{A homogeneous catalogue of projected rotational velocities \\ accounting for limb-darkening}

   \author{
      R.~Varas\inst{1}\thanks{Corresponding author: R. Varas, e-mail: \texttt{rvaras@iaa.es}}
      \and 
      G.~Morello\inst{1,2} 
      \and 
      M.~Zechmeister\inst{3}
      \and 
      P.~J.~Amado\inst{1} 
      \and 
      F.~J.~Pozuelos\inst{1}  
      \and 
      J.~A.~Caballero\inst{4} 
      \and
      A.~Claret\inst{1,5}
      \and 
      C.~Cifuentes\inst{4} 
      \and 
      R.~Morales\inst{1} 
      \and
      A.~Quirrenbach\inst{6}
      \and
      A.~Reiners\inst{3}
      \and
      I.~Ribas\inst{7,8}
      \and
      V.~J.~S.~Béjar\inst{9,10}
      \and
      M.~Cort\'es-Contreras\inst{11}
      \and
      A.~P.~Hatzes\inst{12}
      \and 
      Th.~Henning\inst{13}
      \and
      I.~Hermelo\inst{14}
      \and
      H.~L.~Ruh\inst{3}
      \and
      A.~Schweitzer\inst{15}
      \and
      H.~M.~Tabernero\inst{7,8}
      \and
      M.~R.~Zapatero Osorio\inst{4}
      } 
      
   \institute{
      Instituto de Astrof\'isica de Andaluc\'ia, CSIC, Glorieta de la Astronom\'ia s/n, 18008 Granada, Spain
      \and
      INAF- Palermo Astronomical Observatory, Piazza del Parlamento, 1, 90134 Palermo, Italy
      \and
      Institut f\"ur Astrophysik und Geophysik, Georg-August-Universit\"at G\"ottingen, Friedrich-Hund-Platz 1, 37077 G\"ottingen, Germany
      \and
      Centro de Astrobiolog\'ia, CSIC-INTA, Camino Bajo del Castillo s/n, Campus ESAC, 28692 Villanueva de la Ca\~nada, Madrid, Spain
      \and 
      Universidad de Granada, c/ Gran V\'ia de Col\'on 48, 18010 Granada, Spain
      \and
      Landessternwarte, Zentrum f\"ur Astronomie der Universit\"at Heidelberg, K\"onigstuhl 12, 69117 Heidelberg, Germany
      \and
      Institut de Ci\`encies de l'Espai, CSIC, c/ de Can Magrans s/n, Campus UAB, 08193 Bellaterra, Barcelona, Spain
	   \and
	   Institut d'Estudis Espacials de Catalunya, 08860 Castelldefels, Barcelona, Spain
      \and
      Instituto de Astrof\'isica de Canarias, 38205 La Laguna, Tenerife, Spain
      \and
      Departamento de Astrof\'isica, Universidad de La Laguna, 38206 La Laguna, Tenerife, Spain
      \and
      Departamento de F\'isica de la Tierra y Astrof\'isica \& IPARCOS Instituto de F\'isica de Part\'iculas y del Cosmos, Facultad de Ciencias F\'isicas, Universidad Complutense de Madrid, Plaza de Ciencias 1, 28400 Madrid, Spain
      \and
      Th\"uringer Landessternwarte Tautenburg, Sternwarte 5, 07778 Tautenburg, Germany
      \and
      Max-Planck-Institut f\"ur Astronomie, K\"onigstuhl 17, 69117 Heidelberg, Germany
      \and
      Centro Astron\'omico Hispano en Andaluc\'ia, Observatorio Astron\'omico de Calar Alto, Sierra de los Filabres, 04550 G\'ergal, Almer\'ia, Spain
      \and
      Hamburger Sternwarte, Gojenbergsweg 112, 21029 Hamburg, Germany
   } 

   \authorrunning{R. Varas et al.}

   \date{Received 28 January 2026/ Accepted 10 April 2026}

  \abstract
   {Stellar rotation is closely linked to both age and magnetic activity. Through gyrochronology, it provides a means to estimate stellar ages and trace the evolution of planetary systems, and it is also crucial to constrain and correct stellar activity effects for robust exoplanet detection and characterisation. CARMENES is a dual-channel, high-resolution ($\mathcal{R} >$ 80000) spectrograph that has been highly successful in detecting exoplanets around M-dwarf stars using the radial-velocity technique, and it also enables precise measurements of the projected rotational velocity ($v\sin i$) from spectral line broadening.
   We present an oversampled convolution method incorporating a realistic limb-darkening model to determine $v\sin i$ from CARMENES spectra by comparing observed spectra with that of a template star. The advantages over existing methods in the literature have been assessed using high-resolution synthetic spectra spanning effective temperatures of $2500$--$4000$\,K and projected rotational velocities of up to $50$\,km\,s$^{-1}$. 
   Applied to 392 M dwarfs observed with CARMENES, our method yields $v\sin i$ measurements (or upper limits at $2$\,km\,s$^{-1}$) with a median relative uncertainty of 6.8\,\%, substantially smaller than the 15.4\,\% reported in the literature. 
   This work provides the largest uniform catalogue of $v\sin i$ measurements for M dwarfs, including significantly updated values for several targets, along with 36 new targets.
 }   


   \maketitle

\section{Introduction} \label{sec:intro}

M-dwarf stars are the most numerous stellar population in our Galaxy, constituting roughly 75\,$\%$ of stars in the solar neighbourhood (\citeauthor{henry2006} \citeyear{henry2006}, \citeyear{henry2018}). Their small sizes and masses \citep[0.61--0.075\,\(\textup{M}_\odot\);][]{henry2024}, low temperatures \citep[3900--2300\,K;][]{cifuentes2020}, and long lifetimes make them prime targets for the detection of small, temperate exoplanets \citep{bonfils2013, sabotta2021}. Characterising the rotation of M dwarfs informs us about their magnetic activity, internal structure, and age \citep{suarez-mascareno2015,newton2016,diez-alonso2019,shan2024}. These constraints are essential for interpreting the formation and evolution of their planetary systems.

The Calar Alto high-Resolution search for M dwarfs with Exoearths using Near-infrared and optical Echelle Spectrographs (CARMENES\footnote{\url{https://carmenes.caha.es}}) project provides an excellent dataset for improving rotational velocity measurements in M dwarfs. CARMENES is a dual-channel spectrograph (VIS: 520--960\,nm; NIR: 960--1710\,nm) with high spectral resolution ($\mathcal{R}$\,=\,94600 and $\mathcal{R}$\,=\,80400, respectively), mounted on the 3.5\,m telescope at the Calar Alto Observatory (CAHA) in Almer\'ia, Spain \citep{quirrenbach2014, quirrenbach2018}. The survey has delivered high-quality spectra and radial-velocity (RV) time series for nearly 400 nearby M dwarfs \citep{ribas2023}, enabling the discovery of numerous exoplanets as well as detailed studies of stellar rotation, activity, and variability \citep{baroch2020, jeffers2022, fuhrmeister2018, fuhrmeiste2023, shan2024, ruh2024}.

A fundamental measure of stellar rotation is the projected rotational velocity, $v \sin i$, which represents the component of a star's equatorial rotation along the line of sight. It is typically inferred from the Doppler broadening of absorption lines in high-resolution spectra. Determining $v \sin i$ accurately is challenging, as rotational broadening must be disentangled from other line-broadening mechanisms, such as instrumental effects and stellar turbulence. Conventional approaches often approximate the rotational profile using a convolution kernel that accounts for limb-darkening. However, most studies adopt a fixed linear limb-darkening coefficient (usually $u$\,=\,$0.6$), independent of spectral type or wavelength \citep{jenkins2009, reiners2012, reiners2022}. As demonstrated in this work, this simplification can lead to systematic biases.

Our approach introduces two key enhancements to modelling rotational broadening. First, we apply an oversampled convolution to minimise numerical artefacts. Second, we adopt order-specific limb-darkened kernels that account for wavelength- and temperature-dependent linear coefficients. For each target and spectral order, we compute average $u$ values from stellar-atmosphere models, achieving higher fidelity across the full wavelength range. Tests based on synthetic spectra indicate that this method improves both the precision and accuracy of the results, while remaining computationally efficient.

We apply this method to 392 M dwarfs with CARMENES-VIS observations, enabling direct comparison with previous studies based solely on VIS data. This application yields a homogeneous catalogue\footnote{\url{https://github.com/rvarasg/vsini-limb-darkening}} of $v \sin i$ measurements.
In Appendix \ref{appendix-nir}, we discuss the prospects for extending the method to CARMENES-NIR data.

\section{Observations and data reduction} \label{sec:data}

\subsection{CARMENES spectra}

The CARMENES spectra are reduced using the \texttt{caracal} (CARMENES Reduction And CALibration) pipeline, as detailed by \citet{caballero2016}. Starting with the raw data, \texttt{caracal} performs dark and bias correction, order tracing, flat-relative optimal extraction, and wavelength calibration, resulting in fully reduced and wavelength-calibrated 1-D spectra.

The second pipeline is \texttt{serval} (SpEctrum Radial Velocity AnaLyser, \citeauthor{zechmeister2018} \citeyear{zechmeister2018}). It creates a high signal-to-noise ratio spectrum using the reduced spectra of the target star. It then computes the series of RVs via least-square fitting, telluric masking, \'echelle order weighting, and correcting for systematic effects. The output also includes activity indices. Here we present the implementation in \texttt{serval} of the computation of the $v \sin i$ accounting for the effect of limb-darkening.

\subsection{Synthetic data}

We adopted four spectra from the \texttt{PHOENIX} NewEra stellar atmosphere models \citep{hauschildt2025} and corresponding surface
intensity distributions \citep{claret2025}. The synthetic templates were selected among main-sequence stars with solar metallicity, covering the range $T_{\mathrm{eff}}\,=\,2500$--$4000$\,K in steps of 500\,K. Each template provides intensity spectra at 127\,$\mu$ values, where $\mu$ is the cosine of the angle between the normal to the stellar surface and the line of sight, spanning the wavelength range 1000--60000\,{\AA} with a constant step of 0.05\,{\AA}. These model spectra fully cover the CARMENES wavelength range at slightly higher sampling rate than the instrument.

We numerically integrated the stellar spectrum over a grid of 3000\,$\times$\,3000 square cells covering the sky-projected stellar disc, following the method described by \cite{morello2022} and \cite{canocchi2024}. 
Each cell is identified by the Cartesian coordinates of its centre $(x_j;\,y_j)$, which are normalized to the stellar radius. The corresponding $\mu_j$ values are calculated as:
\begin{equation}
\mu_j = \sqrt{1 - x_j^2 - y_j^2}.
\end{equation}
The ``static'' cell spectrum is then obtained by interpolation at $\mu_j$ from the template, here using a cubic spline, though the method is only weakly sensitive to the choice of interpolation scheme. To account for stellar rotation, we Doppler-shifted the spectrum of each cell according to its velocity along the line of sight,
\begin{equation}
v_j = ( y_j \sin\bar{\lambda} - x_j \cos\bar{\lambda} )\cdot v \sin i,
\end{equation}
where $\bar\lambda$ is a dummy parameter in this context (we set $\bar\lambda$\,=\,0). In the presence of a transiting planet, $\bar\lambda$ traditionally denotes the sky-projected spin-orbit angle. We resampled the Doppler-shifted spectra onto the same wavelength grid as the template through linear interpolation, allowing for the direct summation of the cell spectra.

\section{Methods}\label{sec:methods}

The rotational broadening of stellar spectra can be implemented using several approaches. The two most common ones are convolution with a rotational broadening kernel \citep[]{gray2005} and direct numerical integration \citep[]{carvalho23}. Although numerical integration provides a more general and potentially more accurate procedure, it is computationally more expensive. These methods are often adopted with a single linear limb-darkening coefficient (commonly $u = 0.6$), despite the fact that $u$ depends on both stellar temperature and wavelength. In this work, we develop an oversampled convolution method that offers a fast and robust alternative while properly accounting for wavelength- and temperature-dependent limb-darkening.

The following subsections describe the method and the incremental improvements, tested using synthetic spectra, that guided its development. The test setup consists of a disc-integrated reference spectrum with $v \sin i = 0$\,km\,s$^{-1}$ and a rotationally broadened target spectrum with $v \sin i$ values of 4, 8, 14, 20, 30, and 50\,km\,s$^{-1}$, and effective temperatures $T_{\mathrm{eff}} = 2500, 3000, 3500,$ and 4000\,K. 
These values are representative of the M-dwarf sample observed by CARMENES, although the method itself does not, in principle, impose any restriction on the stellar parameters.
For each case, $v \sin i$ was computed using the linear limb-darkening law with theoretical $u$ coefficients.

\subsection{Classic convolution and grid-based $v \sin i$ fitting}\label{sec:method_convolution}

The standard approach to determining $v \sin i$ involves convolving a chosen template spectrum with a rotational broadening kernel over a grid of trial $v \sin i$ values, and minimising the residuals with respect to the target spectrum.
In this work, we adopt the commonly used semi-circular kernel, which represents a rigidly rotating, uniformly emitting stellar disc:

\begin{equation}
\label{eqn:basic_kernel}
\tilde K(\tilde{v}) = \sqrt{1 - \tilde{v}^2},
\end{equation}
where $\tilde{v}$ is the velocity normalised to $v \sin i$. The kernel is subsequently normalised to conserve flux.

A set of broadened templates is generated by convolving the reference spectrum with the kernel over a grid of $v \sin i$ values with steps of $\Delta (v \sin i) = 0.01$\,km\,s$^{-1}$.
Both the target and template spectra are cubic-interpolated onto the CARMENES wavelength grid, which has an approximately constant spectral resolution and therefore corresponds to a uniform sampling in $\ln \lambda$. Each broadened template is compared with the target spectrum, and the optimal $v \sin i$ is obtained through $\chi^2$ minimization, defined as:
\begin{equation}
    \chi^2(v\sin i) = \sum_i \left(\frac{f_{i}\,-\,f_{\mathrm{tpl}}(\lambda_i, v\sin i)}{\sigma_{i}}\right)^2,
\end{equation}
where $f_{i}$ and $f_{\mathrm{tpl}}(\lambda_i, v\sin i)$ are the target and the artificially broadened template fluxes, respectively, and $\sigma_i$ is the target flux uncertainty. A parabola is then fitted to the $\chi^2$ values at the minimum and its two adjacent grid points, and the vertex is adopted as the final $v \sin i$.
This procedure is applied independently to each corresponding CARMENES-VIS \'echelle order (108--66, 562.3--934.3\,nm). The final $v \sin i$ for a given target is taken as the median of the order-by-order estimates, while the standard deviation provides the uncertainty.

\begin{figure}[]
\includegraphics[width=0.99\hsize]{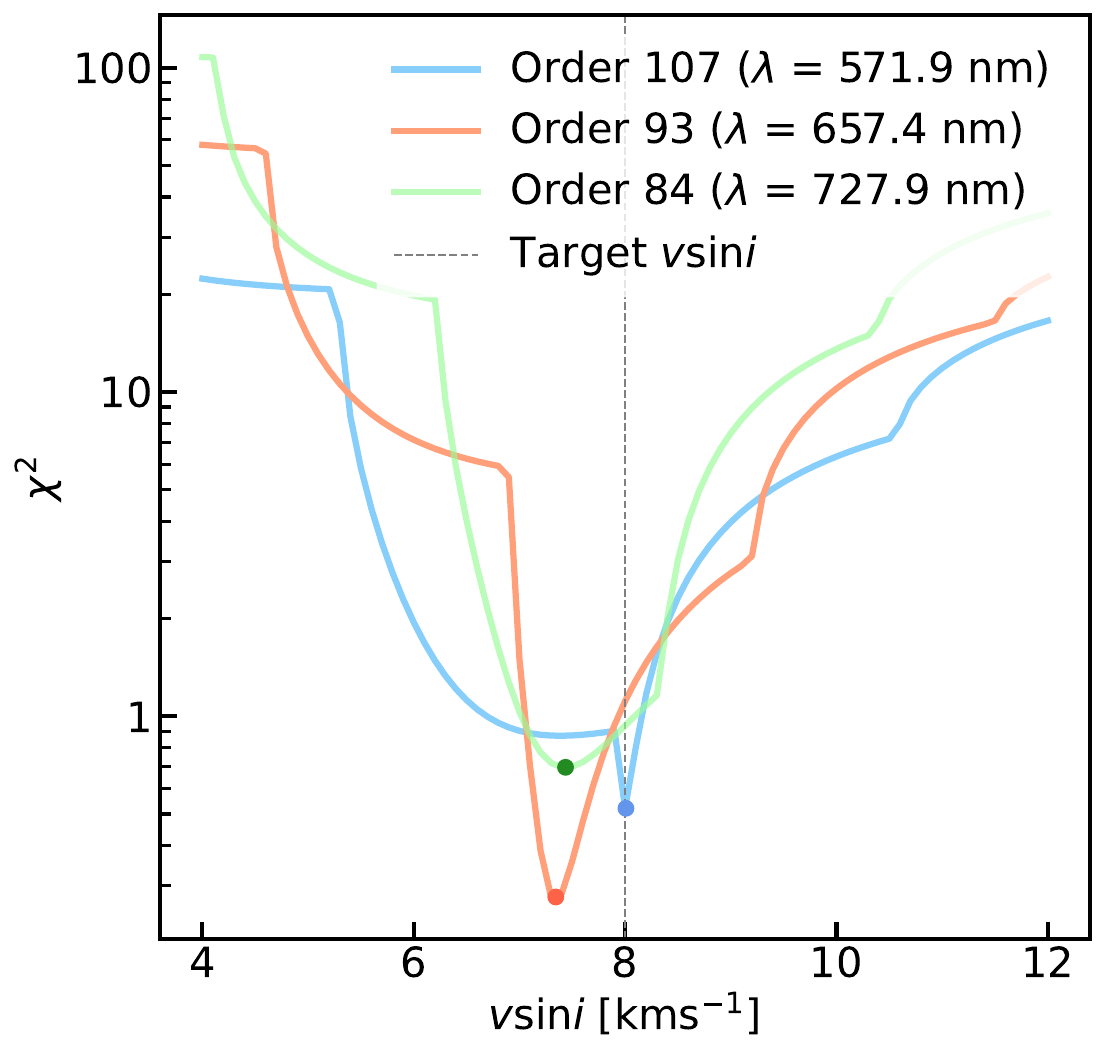}
\caption{$\chi^2$ minimization between reference and target synthetic spectra as a function of $v \sin i$ for three spectral orders, computed without limb-darkening correction. The spectra correspond to $T_\mathrm{eff}$\,=\,3000\,K, with a target value of $v \sin i =8$\,km\,s$^{-1}$ (vertical line).
\label{fig:fig1}}
\end{figure}

Fig.\,\ref{fig:fig1} illustrates the $\chi^2$ minimization as a function of $v \sin i$ for three spectral orders. The resulting $v \sin i$ is indicated by a circular marker. These curves exhibit kinks at different $v \sin i$ values, which bias the inferred velocity depending on their proximity to the true solution. The origin of these kinks is explained in the following subsection.

\subsection{Oversampling}

The numerical discretisation of the kernel in Equation \ref{eqn:basic_kernel} naturally leads to small kinks when varying $v\sin i$, because the kernel is evaluated on a discrete logarithmic wavelength grid. The velocity sampling is fixed by $\Delta v = \Delta(\ln\lambda)\,c$, so that the theoretical kernel half-width is:
\begin{equation}
k_{\max} = \frac{v}{\Delta v}.
\end{equation}
In practice the convolution uses the integer quantity:
\begin{equation}
k = \lfloor k_{\max}\rfloor ,
\end{equation}
which is the number of grid points effectively included in the discrete kernel. The discretised rotational kernel is then computed as:
\begin{equation}
\tilde K(j) = \sqrt{1-\left(\frac{j}{k_{\max}}\right)^{2}}, \qquad j=-k,\ldots,k ,
\end{equation}
and the convolution is performed on the same logarithmic wavelength grid. Each time $v\sin i$ increases enough for $k_{\max}$ to cross an integer boundary, the kernel jumps from $k(v_0)$ to $k(v_1) = k(v_0)+1$, producing a step-wise change that propagates into the $\chi^2$ curve and may shift the position of the minimum.

To mitigate this effect, we oversample the template spectrum before the convolution. It is interpolated (first-order spline) onto a finer grid, convolved with the rotational kernel for the desired $v\sin i$, and then downsampled back to the native wavelength points for the computation of residuals. This approach preserves the original sampling while smoothing the discontinuities introduced by the discrete kernel.The oversampling factor scales linearly with the wavelength grid density, hence, with a factor of $N$ the wavelength grid contains $N$-times more sampling points. Note that factor\,=\,1 means no oversampling is applied.

\begin{figure}[]
\includegraphics[width=0.99\hsize]{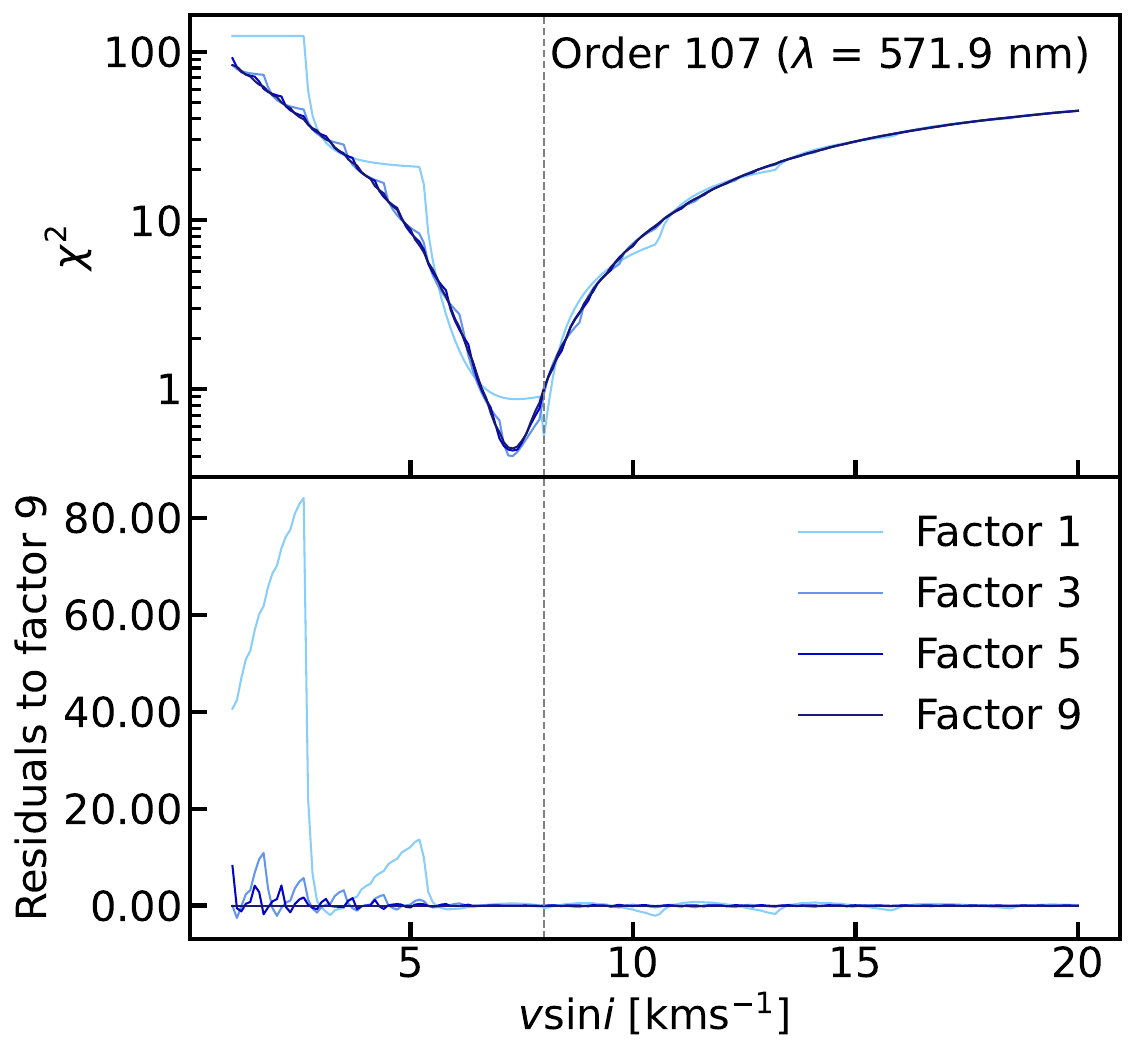}
\caption{Top: $\chi^2$ minimization as a function of $v \sin i$ for different oversampling factors. Bottom: residuals relative to the best-fit solution (oversampling factor\,=\,9). The convolution kernel does not include any limb-darkening effect. The vertical line marks the target value $v \sin i =8$\,km\,s$^{-1}$, obtained from synthetic spectra with $T_\mathrm{eff}$\,=\,3000\,K.  
\label{fig:fig2}}
\end{figure}

\begin{figure}[]
\includegraphics[width=0.99\hsize]{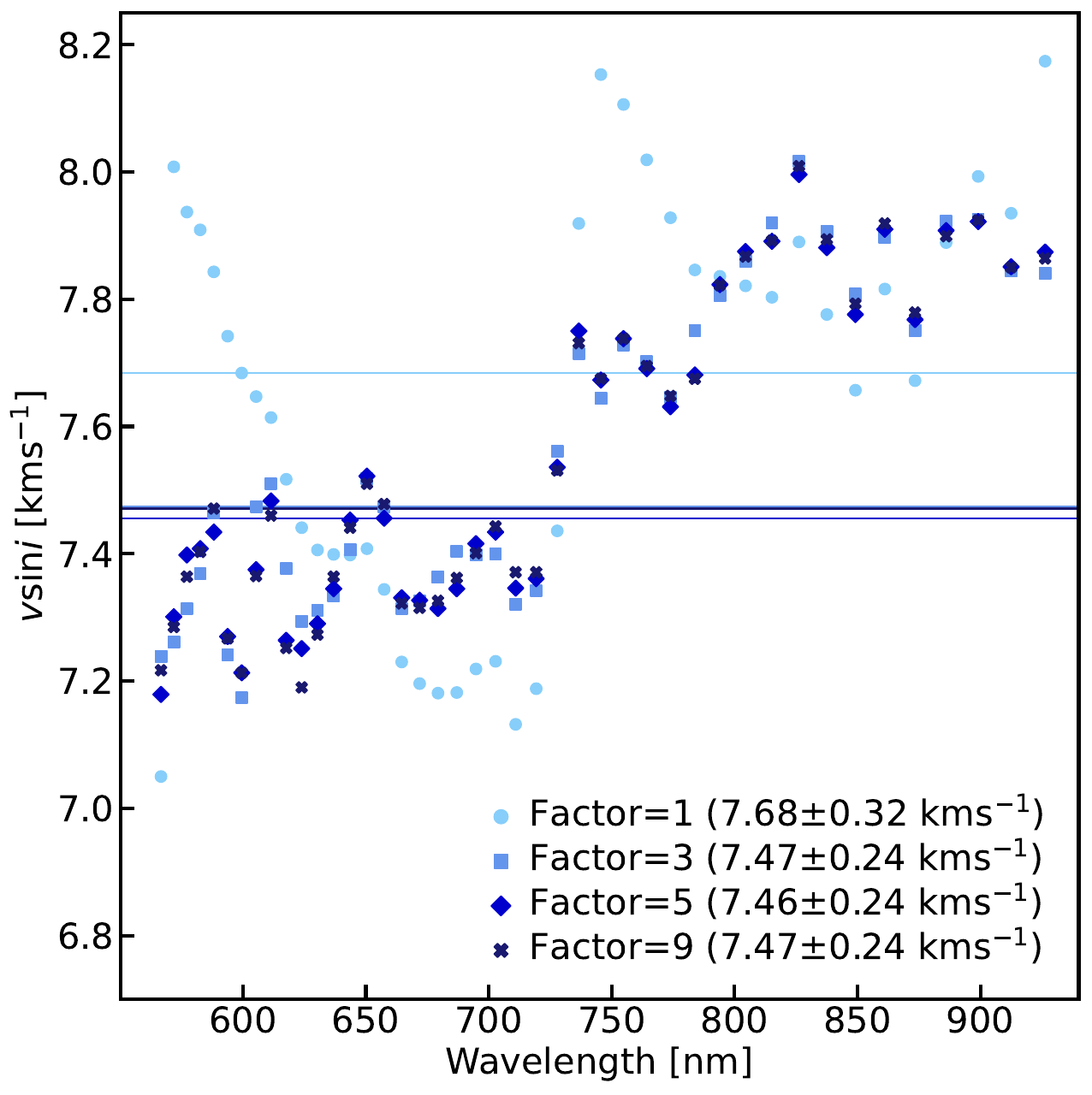}
\caption{Best-fitting $v \sin i$ values as a function of wavelength for different oversampling factors, obtained without accounting for limb-darkening in the convolution process. Horizontal lines correspond to the median values. The spectra correspond to $T_\mathrm{eff}$\,=\,3000\,K and a target value of $v \sin i =8$\,km\,s$^{-1}$.
\label{fig:vsini_factors}}
\end{figure}

Fig.\,\ref{fig:fig2} illustrates how increasing the oversampling factor reduces the amplitude of the discontinuities. Fig.\,\ref{fig:vsini_factors} shows the resulting $v\sin i$ estimates for each spectral order. Oversampling substantially improves the consistency of $v\sin i$ estimates across wavelength. Increasing the factor from 1 to 3 reduces the scatter by $\sim$\,50\,\%. Larger factors do not yield any significant additional gain, with median values and uncertainties essentially unchanged. We therefore adopt an oversampling factor of 5 as an efficient compromise between computational cost and precision.

\subsection{Limb-darkening}
\label{sect:limb_darkening}

We hypothesise that the remaining discrepancies in our analysis (offsets and trends) may be related to stellar limb-darkening, which modifies the rotational broadening kernel through the non-uniform surface brightness of the stellar disc. 
The general form of the limb-darkened rotational kernel is given in Appendix\,\ref{appendix-kernel} (Equation \ref{eqn-integral}).

For a circularly symmetric disc, $\mu = \sqrt{1 - r^2}$, where $r$ is the projected radial coordinate normalised to a reference radius. Several functional forms have been proposed to approximate $I_\lambda(\mu)$; the mathematical derivation of the kernels, including the effect of limb-darkening, is described in Appendix\,\ref{appendix-kernel}. In this work, we find that the linear law \citep{milne1921} provides sufficient precision. More complex prescriptions, such as the power-2 law \citep{hestroffer1997,morello2017}, yield improvements below 0.5\,\% (see Appendix\,\ref{appendix-aspects}).  

The normalised kernel for a linear limb-darkening law in Equation \ref{eqn:kernel-linear} can be written in a discrete form as:
\begin{equation}
\label{eqn:discrete_kernel}
   \tilde K(x) = \frac{2 (1-u) \sqrt{1-x^2} + \frac{\pi}{2} u (1-x^2)}{\pi(1 - \frac{u}{3})},
\end{equation}
where \(x = j / k_{\mathrm{max}}\) and $u$ is the limb-darkening coefficient.

However, because limb-darkening varies with wavelength, a single coefficient cannot reproduce the stellar intensity profile accurately across all spectral orders. 
To account for this wavelength dependence, we determined an optimal coefficient for each spectral order. The simplest approach consists of computing passband-integrated coefficients per order, following standard practice in low-resolution transit spectroscopy or photometry \citep{howarth2011}. We derived these theoretical coefficients using the \texttt{ExoTETHyS} software \citep{morello2020a,morello2020b}, employing the same \texttt{PHOENIX} NewEra stellar templates adopted in our study. 

We also investigated whether photometric limb-darkening coefficients are suboptimal for this application, as they neglect variations between continuum and spectral lines. Therefore, we attempted simultaneous fitting of $u$ and $v \sin i$. However, this approach proved to be computationally expensive and was critically affected by parameter degeneracies (see Appendix\,\ref{appendix-aspects}).

\begin{figure}[]
\includegraphics[width=0.99\hsize]{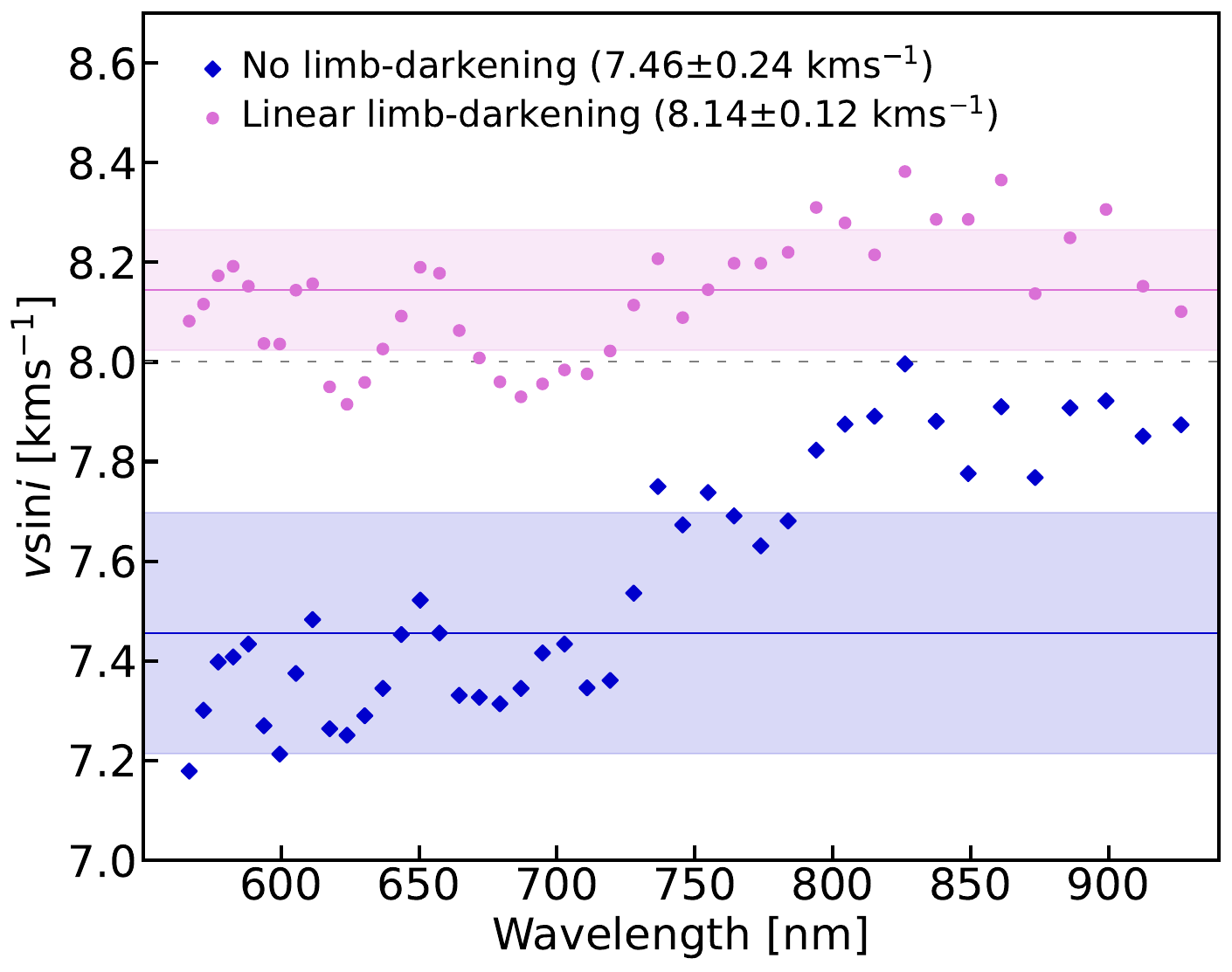}
\caption{Best-fitting $v \sin i$ values as a function of wavelength, obtained with and without accounting for limb-darkening during the convolution. Horizontal solid lines indicate the median values, and the shaded regions represent the standard deviations. The dashed line marks the target value $v \sin i =8$\,km\,s$^{-1}$, for spectra with $T_\mathrm{eff}$\,=\,3000\,K. 
\label{fig:results-ld}}
\end{figure}

Including limb-darkening clearly improves both accuracy and precision, reducing the mean offset and the scatter across spectral orders (Fig.\,\ref{fig:results-ld}). The deviation from the target is reduced by 5--10\,\% (consistent with the decrease in the kernel’s effective width when including limb-darkening; see Appendix\,\ref{appendix-aspects}), and the uncertainty is nearly halved. Top panel of Fig.\,\ref{fig:vsini_teff} provides an overview of the derived $v\sin i$ values across the full range of $T_{\mathrm{eff}}$. Because differences between the tested limb-darkening prescriptions remain below 0.5\,\%, the remaining discrepancies likely originate from the convolution procedure rather than the specific choice of law. Nevertheless, including oversampling and limb-darkening consistently improves the precision of $v\sin i$ (see Appendix\,\ref{sec:ld-oversampling}), motivating our adoption of the linear law.

\subsection{Numerical integration}
\label{sect:subsection_ni}
\citet{carvalho23} developed a method to broaden spectra due to projected rotation, incorporating linear limb-darkening via direct numerical integration. We find that this approach is more accurate and precise than convolution, but at a substantially higher computational cost.

\begin{figure}[]
\centering
\includegraphics[width=0.99\hsize]{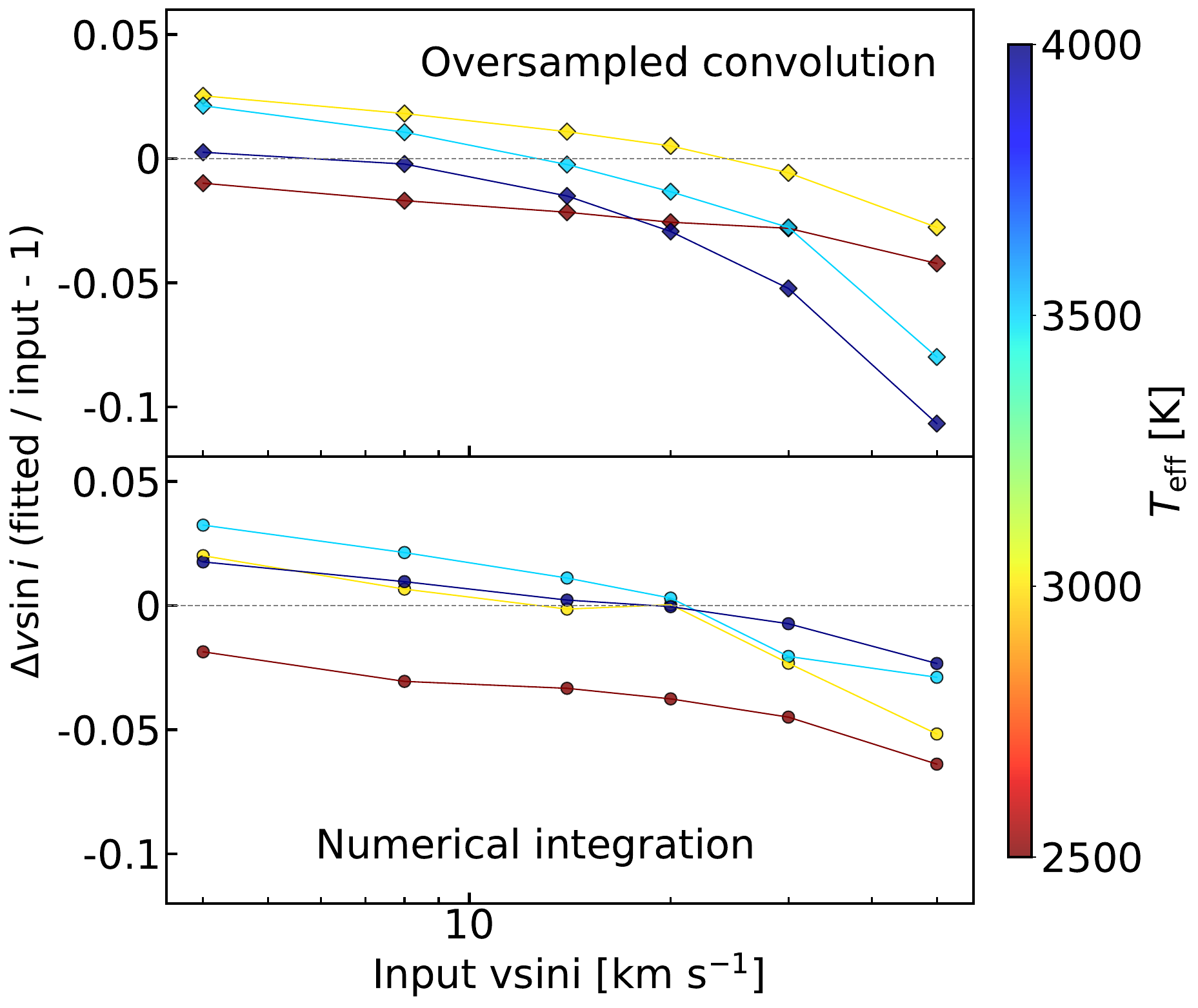}
\caption{Normalized difference between input and fitted $v \sin i$ as a function of input $v \sin i$ and spectra's $T_{\mathrm{eff}}$, for oversampled convolution (top panel) and numerical integration (bottom panel).
\label{fig:vsini_teff}}
\end{figure}

The top panel of Fig.\,\ref{fig:vsini_teff} shows the derived normalized $v \sin i$ as a function of the input values for different $T_{\mathrm{eff}}$ using the oversampled convolution. At high $v \sin i$ ($\gtrsim 20$\,km\,s$^{-1}$), the fitted values increasingly deviate from the input, with the effect more pronounced at higher $T_{\mathrm{eff}}$, particularly at 4000\,K.

The bottom panel of Fig.\,\ref{fig:vsini_teff} shows the results obtained using numerical integration. This method is particularly more precise for hotter ($T_{\mathrm{eff}}$\,=\,$3500$--4000\,K) and faster-rotating ($v \sin i \gtrsim 30$\,km\,s$^{-1}$) M dwarfs, improving precision by up to 5--10\,$\%$. We recommend numerical integration for individual, star-by-star analyses, whereas convolution remains preferable for statistical studies of large stellar samples.

\subsection{Implementation in \texttt{serval} with CARMENES spectra} \label{sec:serval_vsini}

The \texttt{serval} pipeline \citep{zechmeister2018} was originally developed to compute RVs for exoplanet searches, but it also provides additional data products, including activity indicators and $v \sin i$. We adapted our method to operate within the \texttt{serval} framework, following a regression-based approach analogous to cross-correlation techniques \citep{delfosse98, browning10, jeffers2018}.

Based on previous lessons (Sect.\,\ref{sec:methods}), we apply an oversampled convolution (factor\,=\,5) with a linear limb-darkening kernel, using coefficients computed for each stellar temperature and spectral order, to broaden and resample the reference spectrum. Each co-added target spectrum is compared to the corresponding template order via $\chi^2$ minimization over a grid of $v \sin i$ values (0--150\,km\,s$^{-1}$, 0.5\,km\,s$^{-1}$ steps), applied independently to each spectral order. For slow rotators ($v \sin i < 4$\,km\,s$^{-1}$), a finer velocity grid (0--6\,km\,s$^{-1}$, 0.05\,km\,s$^{-1}$ steps) is used to improve resolution near 0\,km\,s$^{-1}$. The final $v \sin i$ is taken as the median of the order-by-order estimates, with the standard deviation providing the uncertainty.

\begin{table}[]
\centering
\caption{List of stars used as template for computing $v \sin i$ depending on spectral type \citep{schofer2019}.}
\label{tab:tpls}
\begin{tabular}{llcc}
\hline
\hline
Karmn & GJ identifier & Spectral type & $T_{\mathrm{eff}}$ {[}K{]}\\ \hline
J14257+236W & GJ 548A & M0.0\,V & 3850 \\
J18580+059 & GJ 740 & M0.5\,V & 3800 \\
J18051–030 & GJ 701 & M1.0\,V & 3680 \\
J16254+543 & GJ 625 & M1.5\,V & 3600 \\
J06103+821 & GJ 226 & M2.0\,V & 3550 \\
J17198+417 & GJ 671 & M2.5\,V & 3450 \\
J15194–077 & GJ 581 & M3.0\,V & 3400 \\
J17578+046 & GJ 699 & M3.5\,V & 3250 \\
J11477+008 & GJ 447 & M4.0\,V & 3200 \\
J19216+208 & GJ 1235 & M4.5\,V & 3100 \\
J03133+047 & GJ 1057 & M5.0\,V & 3050 \\
J00067–075 & GJ 1002 & M5.5\,V & 3000 \\
J07403–174 & GJ 283B & M6.0\,V & 2800 \\
J02530+168 & GJ 10393 & M6.5--9.0\,V & 2700--2450 \\
\hline
\end{tabular}
\end{table}

Templates are chosen according to the target’s sub-spectral type (Table\,\ref{tab:tpls}), following \citet{schofer2019}, who identified the least active and slowest rotator for each subtype. For M6.5\,V and later, Teegarden’s star (J02530+168) is adopted as reference. All templates are generated with \texttt{serval}, ensuring a consistent instrumental profile across all spectral orders. The numerical integration method by \citet{carvalho23} has been implemented as an alternative alongside convolution. 

\section{Results} \label{sec:results} 

Using CARMENES-VIS spectra of 392 M dwarfs, we present a catalogue of projected rotational velocities that incorporates limb-darkening effects (Table\,\ref{tab:vsini_results}). The CARMENES spectra are intrinsically broadened by the instrument, corresponding to an effective $v \sin i = 2$\,km\,s$^{-1}$ \citep{reiners2018}. Consequently, only an upper limit of 2\,km\,s$^{-1}$ can be placed on stars rotating more slowly than this threshold. Extension to the NIR is deferred to future work; see Appendix\,\ref{appendix-nir} for a preliminary assessment of its feasibility and potential advantages.

To balance accuracy and computational efficiency, we do not always use all available spectra for each target. For stars with fewer than 20 spectra, all are included; otherwise, we select 20 evenly spaced spectra. This selection follows the behaviour of the underlying \texttt{serval}-inherited function. We have verified that using more than 20 spectra does not significantly improve the precision of the derived $v \sin i$.

The command-line input files used to run \texttt{serval} are provided in our GitHub repository. Template stars have projected rotational velocities below the instrumental resolution and thus have $v \sin i < 2$\,km\,s$^{-1}$ (see Table\,\ref{tab:tpls}). In some cases, adjustments were necessary, for example, for certain M3.5\,V stars it is preferable to use J14310-122 as the template rather than J17578+046. For stars later than M6.0\,V, only the reddest orders (orders 88--66, 688.8--934.3\,nm) are used due to low signal-to-noise in the bluer orders.

Theoretical linear limb-darkening coefficients were computed for each spectral order at $T_{\mathrm{eff}} = 2500, 3000, 3500,$ and 4000\,K and assuming solar metallicity. These coefficients are available in the associated GitHub repository. For intermediate temperatures, the coefficients are computed via linear interpolation in \texttt{serval}. 

The stellar temperatures are obtained from Carmencita \citep{caballero2017}, that gathers values from several works \citep{soto2020, palle2020,cifuentes2020,marfil2021}.
Carmencita is the "CARMEN(ES) Cool dwarf Information and daTa Archive", and contains a large number of parameters for more than 2000 M dwarfs in the solar neighbourhood brighter than J = 11.5 mag.

\begin{table*}[]
\centering
\caption{Catalogue of $v \sin i$ values for 392 M-dwarf stars of the CARMENES sample using VIS spectra.\,$^a$ }
\label{tab:vsini_results}
\begin{tabular}{lccccccc}
\hline
\hline
Karmn & Name & Adopted $v \sin i$ & $v \sin i$ OC & $v \sin i$ NI & $v \sin i$ literature & Ref.\,$^b$ & $v_\mathrm{eq}$     \\ 
 & & {[}km\,s$^{-1}${]} & {[}km\,s$^{-1}${]} & {[}km\,s$^{-1}${]} & {[}km\,s$^{-1}${]} & & {[}km\,s$^{-1}${]}\\ \hline
J00051+457 & BD+44 4548 & < 2 & 2 & 2 & 2 & Rein18 & 1.61\,$\pm$\,0.01 \\
J01019+541 & G 218-020 & 29.3\,$\pm$\,0.9 & 29.3\,$\pm$\,0.92 & 29.2\,$\pm$\,0.9 & 30.6\,$\pm$\,3.06 & Rein18 & 28.2\,$\pm$\,0.26 \\
J01056+284 & GJ 1029 & < 2 & 2 & 2 & 4.1\,$\pm$\,0.41 & Jen09 & 0.65\,$\pm$\,0.02 \\
J02002+130 & TZ Ari & < 2 & 2 & 2 & 2 & Rein18 & 4.04\,$\pm$\,0.08 \\
J02164+135 & LP 469-206 & 13.4\,$\pm$\,0.7 & 13.4\,$\pm$\,0.7 & 13.4\,$\pm$\,0.7 & 12.4\,$\pm$\,1.24 & Barn14 & ... \\\hline
\end{tabular}
\begin{minipage}{\hsize}
\vspace{0.7ex}
\textbf{Notes.} $^{(a)}$\,Only five rows are shown as an example, the complete table is available in the GitHub repository. Listed are: star's Karmn identifier, most common or discovery name, adopted $v \sin i$, $v \sin i$ obtained using oversampled convolution (OC), numerical integration (NI), the value from the literature and reference to the work, and the equatorial rotation velocity. $^{(b)}$
Jen09: \citet{jenkins2009}, 
Barn14: \citet{barnes2014},
Rein18: \citet{reiners2018}.
\end{minipage}
\end{table*}

\begin{figure}[]
\centering
\includegraphics[width=0.99\hsize]{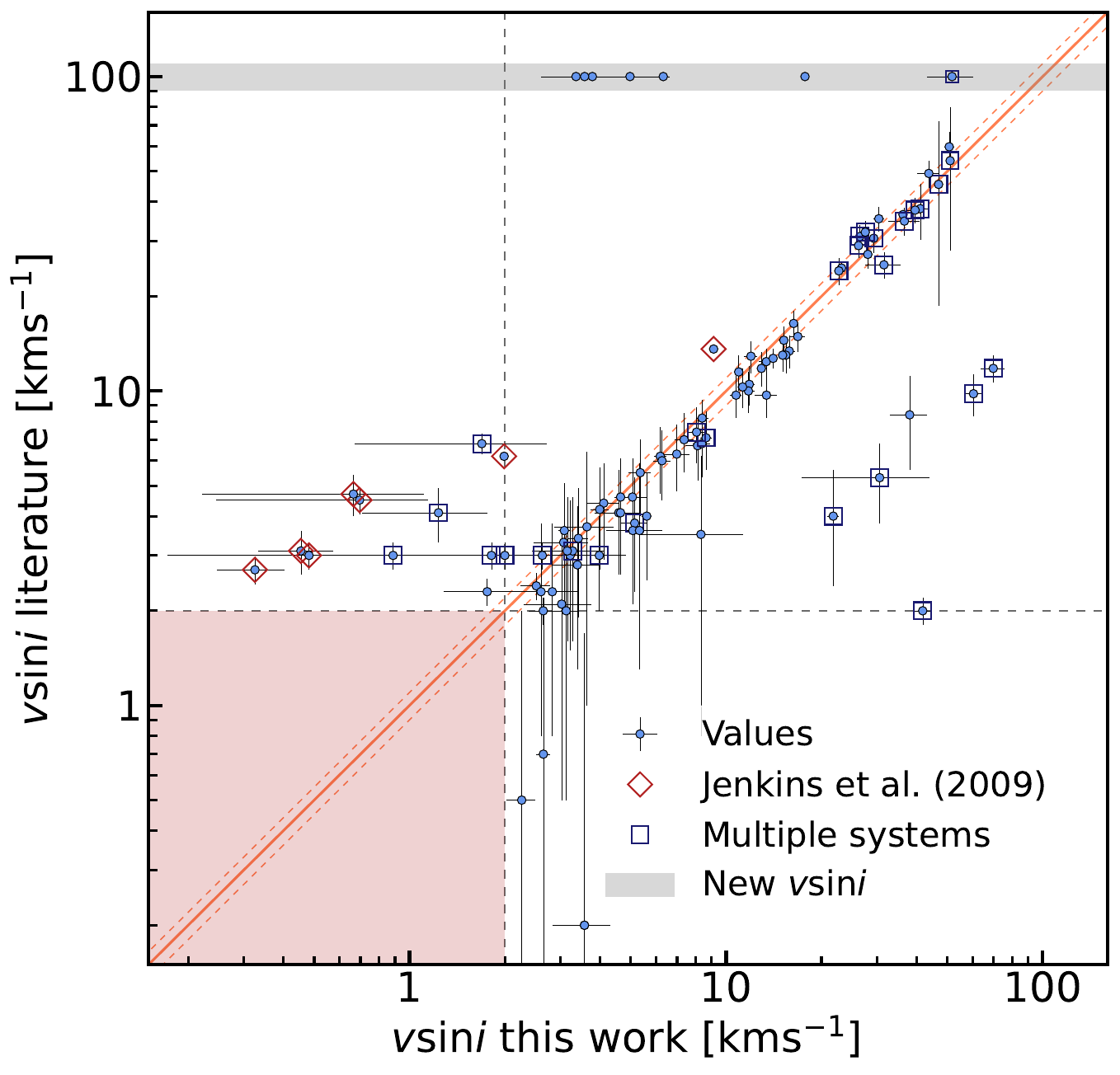}
\caption{$v \sin i$ values obtained in this work using the oversampled convolution method compared with literature measurements. Newly determined $v \sin i$ values not previously available are shown in the shaded area at $v \sin i$ (literature)\,=\,100\,km\,s$^{-1}$. The solid line indicates the 1:1 relation, while the dashed lines correspond to a 10\,$\%$ deviation. Grey dashed lines and the shaded area mark the 2\,km\,s$^{-1}$ limit of CARMENES. Values inside this region are not plotted.
\label{fig:vsini}}
\end{figure}

\subsection{Oversampled convolution}

We compare the 392 $v \sin i$ values derived in this work with literature measurements compiled in Carmencita, adopting updated values from \citet{reiners2022} where available (see Table\,\ref{tab:vsini_results}). The adopted $v \sin i$ is the value we found to be the most reliable one among our results, using oversampled convolution (OC) and numerical integration (NI), the literature, and the cross check with the equatorial velocity ($v_\mathrm{eq}$). A comparison with other works \citep{passegger2020,mas-buitrago2024} is discussed in Appendix\,\ref{appendix-comparison}.

Most of the derived $v\sin i$ values lie below the instrumental resolution limit of CARMENES (2\,km\,s$^{-1}$), in agreement with their literature values (see Table\,\ref{tab:vsini_results}). The exiting outliers can be seen in Fig.\,\ref{fig:vsini}; those points deviating from the 1:1 diagonal line.
Our $v \sin i$ measurements are generally in good agreement with the literature also above 2\,km\,s$^{-1}$, with a small number of outliers. The discrepancies are mainly due to either underestimated or overestimated values in previous studies, or to unresolved multiple systems affecting the observed spectra, which make the derived $v \sin i$ unreliable.

Close to the resolution limit, we find some discrepancies with values reported in the literature. Three of our targets (J01339-176, J04376+528, and J09144+526) show $v \sin i$ measurements very near the $2\,\mathrm{km\,s^{-1}}$ limit, differing from that threshold by less than $0.25\,\mathrm{km\,s^{-1}}$. Of the remaining three objects, J05365+113 and J19346+045 exhibit $v \sin i > 2 \mathrm{\,km\,s^{-1}} > v_\mathrm{eq}$, so only upper limits can be assigned. For J13450+176, our value agrees with that reported by \citet{reiners2018}, but not with the measurement of \citet{reiners2022}.

\begin{figure}[]
\centering
\includegraphics[width=0.95\hsize]{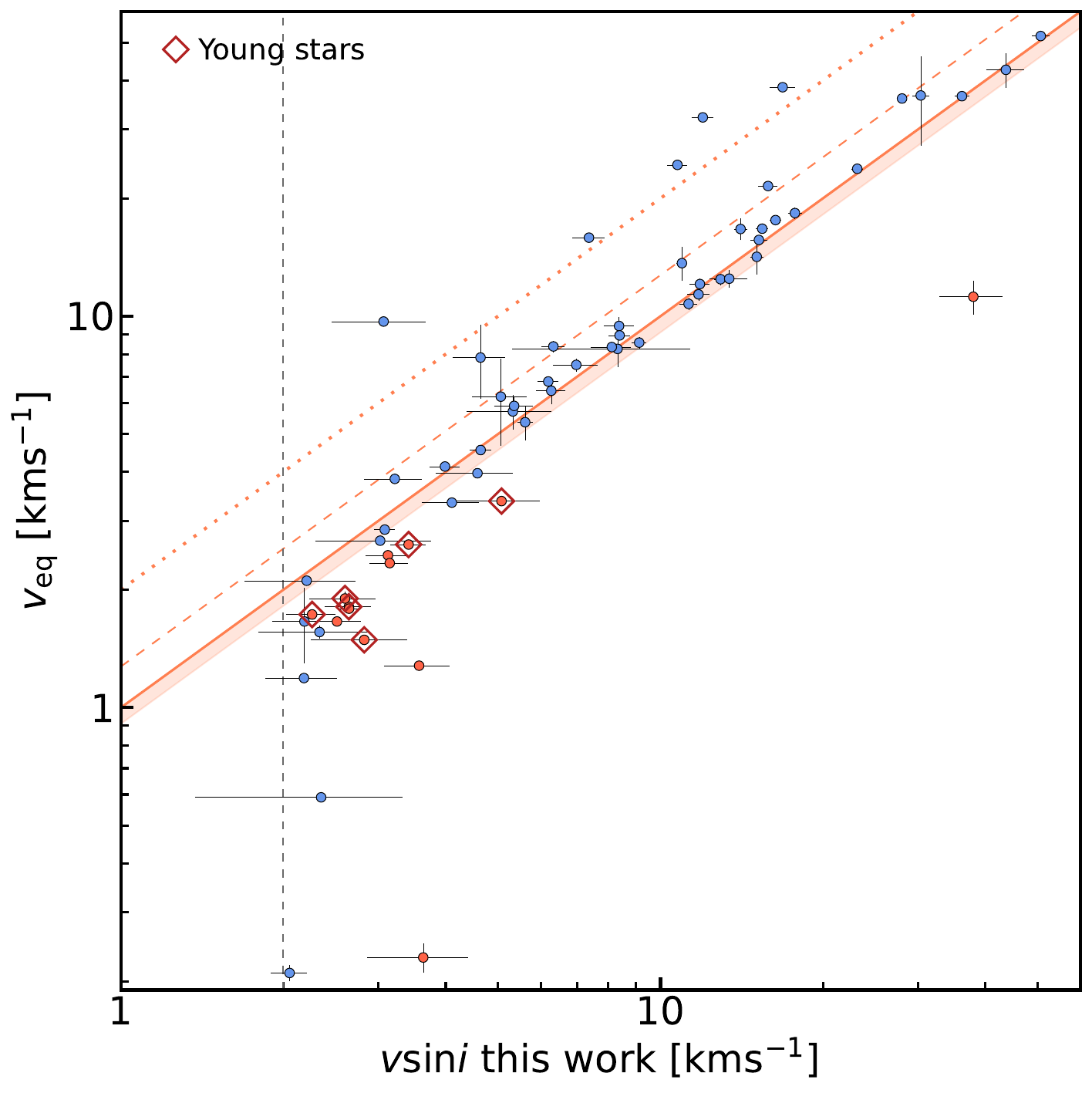}
\caption{Values of $v \sin i$ derived from the oversampled convolution method compared to $v_\mathrm{eq}$. The linear 1:$\tfrac{4}{\pi}$ relation (dashed line) represents the expected value of $\sin i$ for an isotropic distribution of inclination angles (see Appendix B of \citeauthor{ruh2024} \citeyear{ruh2024}). The 1:1 relation with a 10\,\% margin (solid line and shaded region) and the 1:$\tfrac{1}{2}$ relation (dotted line) correspond to $i = 90^\circ$ and $i=30^\circ$, respectively. The vertical grey dashed line indicates the 2\,km\,s$^{-1}$ threshold, and the diamond markers correspond to young candidates \citep{cortescontreras2024}.
\label{fig:vsini_veq}}
\end{figure}

In Fig.\,\ref{fig:vsini_veq}, we compare our $v \sin i$ measurements from the oversampled convolution with limb-darkening to the equatorial velocities $v_\mathrm{eq}$, computed as $v_\mathrm{eq} = 2\pi R_\star / P_\mathrm{rot}$ for stars with available $R_\star$ and $P_\mathrm{rot}$. Errors in the rotational periods were calculated following Equation 4 of \citet{shan2024}, and uncertainties in $v_\mathrm{eq}$ were propagated accordingly. Stars for which $v \sin i > v_\mathrm{eq}$ are highlighted in orange and listed in Table\,\ref{tab:outliers}. 

We revised these outliers and identified six young candidates (J01339-176, J05366+112, J09144+526, J11026+219, J11476+002, J23548+385) as classified in \citet{cortescontreras2024} based on their kinematics and activity indicators (measured from rotation, X rays, ultraviolet and Halpha emission). These targets are highlighted with red diamonds in Fig.\,\ref{fig:vsini_veq}. According to this study, all of them are fast rotators for their spectral types and all but J05366+112 present X rays luminosities and (NUV-J) levels above the defined threshold for active stars. In addition, they present significant Halpha emission with the exception of J09144+526, which has instead lithium measurement \citep{bischoff2020}. In particular, J11476+002 and J23548+385 are candidates of IC2391 and UMa groups, having both of them intense magnetic fields measured in \citet{reiners2022}.

Among the remaining outliers, two of them have a doubtfull youth assignment in the cited work: J04376+528 and J05365+113. Both are kinematically young (Hyades or galactic young disc) and only a slightly faster rotation compared to others. Another one (J19346+045) is a field star with a rotation period of 8.04\,d \citep{diez-alonso2019}, and evidence of chromospheric activity (NUV-J and Halpha emission). Regarding J07361-031 is a SB1 binary \citep{poveda2009} in a quadruple system. Finally, J19346+045, J08536-034 and J19169+051S present no relevant information regarding youth or activity.

We report 36 new $v \sin i$ measurements that were not previously available. In Fig.\,\ref{fig:vsini}, these are represented in the grey-shaded region at $v \sin i_\mathrm{literature} = 100$\,km\,s$^{-1}$. All values appear reliable except for J10182-204, a multiple system where $v \sin i$ is significantly higher than $v_\mathrm{eq}$. Because $v_\mathrm{eq} < 2$\,km\,s$^{-1}$, an upper limit of 2\,km\,s$^{-1}$ can be adopted for this target.

In total, 21 $v \sin i$ measurements in our sample are considered unreliable. Table\,\ref{tab:outliers} summarises the reasons for their unreliability and provides recommended values. For most of these stars, an upper limit can be imposed based on $v_\mathrm{eq}$, while in the remaining cases the literature $v \sin i$ is deemed more reliable and is adopted as the preferred value.
Out of the 33 multiple systems in our sample identified by \citet{cifuentes2025}, 11 show discrepancies between our measurements and literature values. Seven of these are considered unreliable in our work and are listed in Table\,\ref{tab:outliers} as close.

\begin{table}[]
\centering
\caption{Targets with unreliable $v \sin i$ values.}
\label{tab:outliers}
\begin{tabular}{lcll}
\hline
\hline
Karmn & $v \sin i$ & Ref.$^a$ & Comments\,$^b$\\ 
 & [km\,s$^{-1}$] & &  \\ \hline
J00162+198W & $< 2$   & Rein22 & Close \\
J01339-176   & $< 2$ & This work & $v \sin i > v_\mathrm{eq}$ \\
J04376+528  & $< 2$ & This work & $v \sin i > v_\mathrm{eq}$ \\
J05337+019  & 9.8 $\pm$ 1.0  & Rein18 & Close \\
J05365+113  & $< 2$ & Rein22 & $v \sin i > v_\mathrm{eq}$ \\
J05366+112  & $\leq v_\mathrm{eq}$ & This work & $v \sin i > v_\mathrm{eq}$ \\
J05394+406  & 5.3 $\pm$ 0.5 & Rein22 & Close\\
J07033+346  & $< 2$  & This work & $v \sin i > v_\mathrm{eq}$ \\
J07361-031  & $\leq v_\mathrm{eq}$ & This work & Close \\
J08536-034  & 8.4 $\pm$ 0.8 & Rein22 & $v \sin i > v_\mathrm{eq}$\\
J09144+526  & $< 2$ & This work & $v \sin i > v_\mathrm{eq}$ \\
J10182-204  & $< 2$ & This work & $v \sin i > v_\mathrm{eq}$ \\
J11026+219  & $< 2$ & This work & $v \sin i > v_\mathrm{eq}$ \\
J11055+435  & $< 2$ & Rein18 & $v \sin i > v_\mathrm{eq}$ \\
J11476+002  & $< 2$ & This work & $v \sin i > v_\mathrm{eq}$ \\
J16343+571  & 10.22 $\pm$ 1.02 & Mor09  & Close  \\
J19169+051S & $< 2$ & This work & $v \sin i > v_\mathrm{eq}$ \\
J19346+045  & $< 2$ & Rein18 & $v \sin i > v_\mathrm{eq}$ \\
J20198+229  & 11.8 $\pm$ 1.2 & Jen09 & Close \\
J23548+385  & 3.6 $\pm$ 0.4  & Rein18 & $v \sin i > v_\mathrm{eq}$\\
J23585+076  & $< 2$ & Rein18 & Close \\ \hline
\end{tabular}
\begin{minipage}{\hsize}
\vspace{0.7ex}
\textbf{Notes.} $^{(a)}$\,Jen09: \citet{jenkins2009}, Mor09: \citet{morales09}, Rein18: \citet{reiners2018}, Rein22: \citet{reiners2022}. $^{(b)}$\,Comments indicates why they are unreliable. Close systems are identified by \citet{cifuentes2025}.
\end{minipage}
\end{table}

Among the remaining systems, J01056+284, J14155+046, and J15412+759 have $v \sin i > v_\mathrm{eq}$ in the literature, whereas our values are consistent, being below 2\,km\,s$^{-1}$. J05532+242 lacks a measured $v_\mathrm{eq}$, but \citet{jeffers2018} imposed an upper limit of 3\,km\,s$^{-1}$; we adopt 2\,km\,s$^{-1}$. For J14155+046, our measured $v \sin i$ is below 2\,km\,s$^{-1}$, and the spectrum is more consistent with a slow rotator than with the higher literature value of 6.8\,km\,s$^{-1}$ (see Appendix\,\ref{appendix-J14155}).

We identified seven $v \sin i$ measurements from \citet{jenkins2009} that are not in agreement with our results (red diamonds in Fig.\,\ref{fig:vsini}): J02022+103, J02465+164, J03090+100, J14578+566, J15100+193, J15238+174, and J20305+654. For three of these targets, the values from the cited work exceed $v_\mathrm{eq}$, whereas our $v \sin i$ measurements are consistent. The remaining targets have $v \sin i < 2$\,km\,s$^{-1}$ in our analysis and their spectra are better matched by a slowly rotating template than by a broadened one. We therefore recommend adopting our values as updated measurements for these stars.

\begin{figure}[]
\includegraphics[width=0.99\hsize]{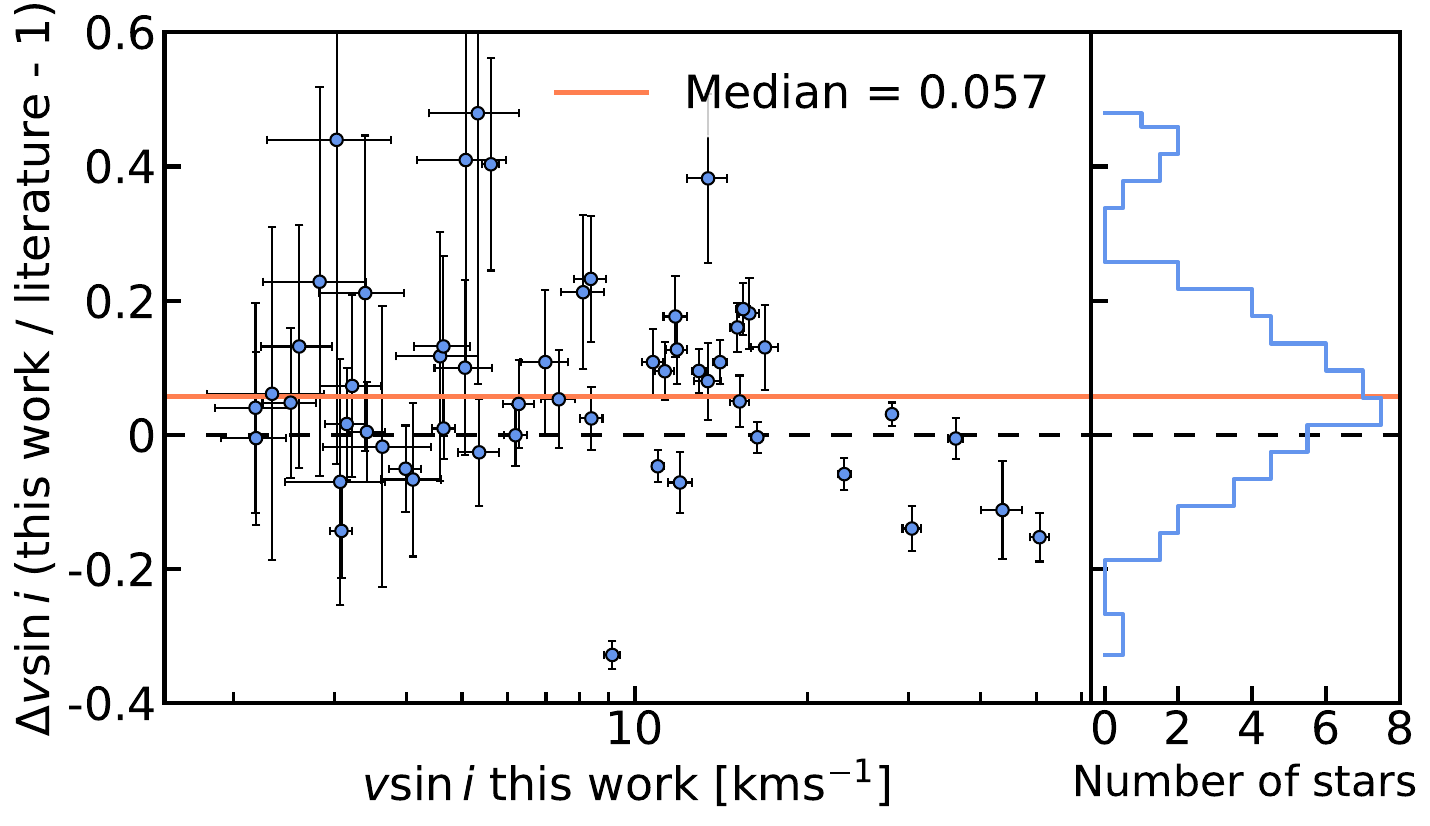}
\caption{Histogram of the normalized difference between the $v \sin i$ value ($>$\,2\,km\,s$^{-1}$) computed with oversampled convolution and the literature, excluding the outliers. The median is marked with a horizontal solid line.
\label{fig:vsini_histogram}}
\end{figure}

Excluding outliers and considering only $v \sin i > 2$\,km\,s$^{-1}$, our measurements are systematically higher than literature values by a median of 5.7\,$\%$, as illustrated in Fig.\,\ref{fig:vsini_histogram}. An increase up to 10\,\% was expected due to the inclusion of limb-darkening (depending on the $T_\mathrm{eff}$ and $v\sin i$; see Sect.\,\ref{sect:limb_darkening} and Appendix\,\ref{appendix-aspects}). At high rotation rates ($v \sin i \gtrsim 20$\,km\,s$^{-1}$), values may be slightly underestimated owing to intrinsic limitations of the oversampled convolution method (see Fig.\,\ref{fig:vsini_teff}). The median relative uncertainty of our $v \sin i$ values above 2\,km\,s$^{-1}$ is 6.8\,\%, compared to 15.4\,\% for the literature.

\subsection{Numerical integration} \label{sec:numerical:integration}

\begin{figure}[]
\includegraphics[width=0.99\hsize]{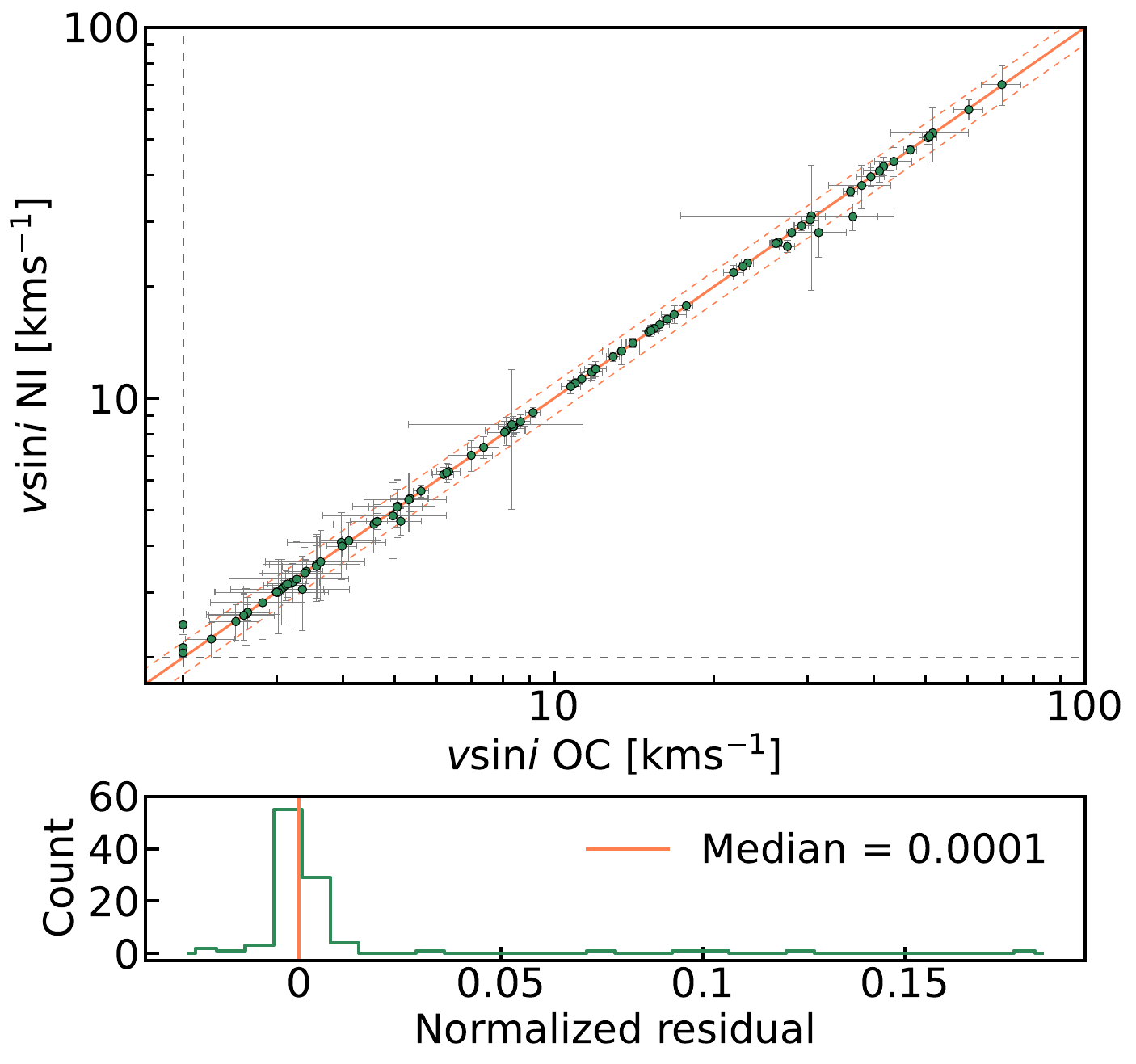}
\caption{Top panel: comparison between $v \sin i$ computed with numerical integration (NI) and oversampled convolution (OC). Dashed lines represent the 2\,km\,s$^{-1}$ limit, and diagonal lines mark the 1:1 relation $\pm$ 10\,$\%$. Bottom panel: histogram of the normalized residuals respect to the 1:1 relation for $v \sin i >$ 2\,km\,s$^{-1}$. Median value represented by a vertical line.
\label{fig:conv_vs_int}}
\end{figure}

The oversampled convolution (OC) method is approximately 3.5 times faster than the numerical integration (NI) approach, although the latter is expected to yield higher precision (see Sect.\,\ref{sect:subsection_ni}).
Table\,\ref{tab:vsini_results} lists all $v \sin i$ values obtained with both methods, while Fig.\,\ref{fig:conv_vs_int} compares their results, showing excellent agreement. The histogram of the normalised residuals displays a median difference of only 0.1\,$\%$, smaller than the offsets predicted from synthetic data in Sect.\,\ref{sect:subsection_ni}, where differences of about 1--2\,$\%$ were found for slow and intermediate rotators, and up to 5--10\,$\%$ for the fastest rotators (see Fig.\,\ref{fig:vsini_teff}).

Overall, values below and above 2\,km\,s$^{-1}$ are fully consistent between methods. The substantial increase in computational time relative to the modest gain in precision does not justify using NI instead of OC, particularly for statistical studies or slow-to-moderate rotators ($v \sin i < 20$\,km\,s$^{-1}$). As discussed in Sect.\,\ref{sect:subsection_ni}, the improvement from NI becomes noticeable only for faster rotators (30--50\,km\,s$^{-1}$) and hotter stars ($T_{\mathrm{eff}} = 4000$\,K). However, this effect remains minor for the CARMENES sample, yielding a similar median error as the OC, 6.5\,\%. 

\section{Conclusions} \label{sec:conclusion}

We have presented a novel method to determine the projected rotational velocity ($v \sin i$) of stars from CARMENES-VIS spectra. This approach upgrades the \texttt{serval} pipeline by (i) adopting an order-dependent rotational broadening kernel that accounts for wavelength- and temperature-dependent stellar limb-darkening, and (ii) performing spectral oversampling prior to convolution with the template spectrum. The effectiveness of these improvements has been tested using both synthetic and real datasets.

Applying this method, we derived a homogeneous catalogue of $v \sin i$ measurements for 392 M dwarfs observed with CARMENES-VIS. Our results show overall good agreement with previously published values, while achieving a substantially higher precision. The median relative uncertainty of our measurements is 6.8\%, compared to 15.4\% for literature values. Additionally, we report 36 new $v \sin i$ measurements that were not previously available.

This catalogue provides a robust and homogeneous reference for studies of stellar rotation in low-mass stars and will be particularly valuable for the characterisation of M-dwarf planetary systems, including applications to age estimates, radial-velocity surveys, and spin–orbit investigations.

\section{Data availability}

Our pipeline, data, and results are available in a \texttt{GitHub} repository at \url{https://github.com/rvarasg/vsini-limb-darkening}. 
Table\,\ref{tab:vsini_results} is available in electronic form at the CDS via anonymous ftp to \url{cdsarc.u-strasbg.fr} (130.79.128.5) or via \url{http://cdsweb.u-strasbg.fr/cgi-bin/qcat?J/A+A/}. 
The CARMENES DR1 dataset, which includes most of the spectra used in this work, is publicly available at \url{https://carmenes.cab.inta-csic.es}.

\begin{acknowledgements}

CARMENES is an instrument at the Centro Astron\'omico Hispano en Andaluc\'ia (CAHA) at Calar Alto (Almer\'{\i}a, Spain), operated jointly by the Junta de Andaluc\'ia and the Instituto de Astrof\'isica de Andaluc\'ia (CSIC).

Funding for CARMENES has been provided by the 
Max-Planck-Institut f\"ur Astronomie (MPIA), 
Consejo Superior de Investigaciones Cient\'{\i}ficas (CSIC),
European Regional Development Fund (ERDF), 
Ministerio de Ciencia, Innovaci\'on y Universidades (MICIU),
Deutsche Forschungsgemeinschaft (DFG),
and the members of the CARMENES Consortium
(\url{https://carmenes.caha.es}).

This publication was based on observations collected under the CARMENES Legacy-Plus project.

The authors wish to express their sincere thanks to all members of the Calar Alto staff for their expert support of the instrument and telescope operation.

We used data from the CARMENES data archive at CAB (CSIC-INTA).

We acknowledge financial support from the Agencia Estatal de Investigaci\'on (AEI/10.13039/501100011033) of the MICIU and the ERDF ``A way of making Europe'' through projects PID2022-137241NB-C4[1:4], PID2021- 125627OB-C3[1:2], PID2019-109522GB-C52, RYC2022-037854-I, and RYC2021-031640-I (2023AT003), and from the Centre of Excellence ``Severo Ochoa'' and ``Mar\'ia de Maeztu'' awards to the Instituto de Astrof\'isica de Andaluc\'ia (CEX2021-001131-S) and Institut de Ci\`encies de l’Espai (CEX2020-001058-M). 
We also acknowledge financial support from the project AST22\_00001\_8 of the Junta de Andaluc\'ia and the Ministerio de Ciencia, Innovaci\'on y Universidades funded by the NextGenerationEU and the Plan de Recuperaci\'on, Transformaci\'on y Resiliencia.

\end{acknowledgements}
\bibliographystyle{aa}
\bibliography{vsini-bib}{}

\begin{appendix}

\section{$v \sin i$ using near-infrared spectra} \label{appendix-nir}
We analysed $v \sin i$ using both CARMENES-NIR and VIS spectra for the coldest stars in the sample (16 objects). This selection is motivated by their higher near-infrared S/N relative to hotter stars, comparable to the VIS S/N \citep{reiners2018}. J07403-174 and J02530+168 were used as templates (Table\,\ref{tab:tpls}), and their $v \sin i$ values are not computed here. For this preliminary analysis, we did not apply oversampling nor include limb-darkening.

\begin{figure}[]
\includegraphics[width=0.99\hsize]{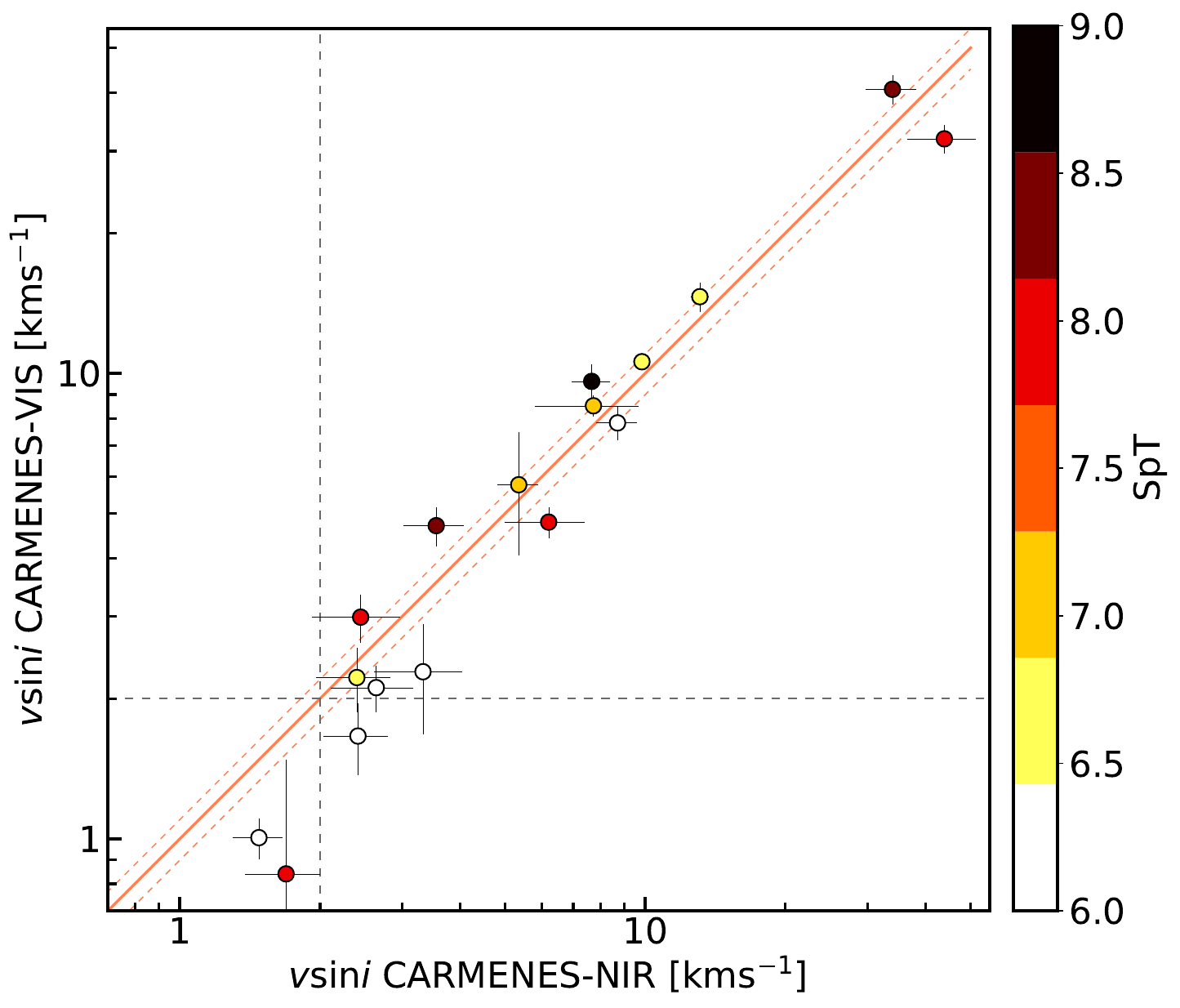}
\caption{Comparison of $v \sin i$ derived from CARMENES-VIS and NIR spectra for M dwarfs later than M6.0\,V. No oversampling or limb-darkening was applied. Lines indicate the 1:1 relation ($\pm$\,10\,$\%$) and the 2\,km\,s$^{-1}$ limit.
\label{fig:vsini-nir}}
\end{figure}

Fig.\,\ref{fig:vsini-nir} presents the comparison of $v \sin i$ derived from NIR and VIS spectra for this subsample. Most measurements are consistent between the two datasets, with a few cases close to or below the 2\,km\,s$^{-1}$ limit. A more detailed analysis, analogous to that performed for the VIS data in this study, is required to fully characterise the small remaining differences between the NIR and VIS results.

\section{Rotation kernel with limb-darkening}
\label{appendix-kernel}

Following \cite{gray2005}, the disc-integrated spectrum of a rigidly rotating, spherical star can be written as a convolution:
\begin{equation}
\label{eqn-integral}
F(v) = \int_{-v\sin i}^{+v\sin i} S(v - v') \, \tilde{K}(v') \, dv',
\end{equation}
where $S(v)$ is the intrinsic, non-rotating stellar spectrum, and 
\begin{equation}
v = c \frac{\lambda - \lambda_{\rm ref}}{\lambda_{\rm ref}}
\end{equation}
is the Doppler velocity relative to a reference wavelength. The kernel $\tilde{K}(v')$ represents the relative contribution of all surface elements sharing the same projected line-of-sight velocity $v'$. Choosing the $x$-axis along the projected stellar equator, the line-of-sight velocity of a surface element is:
\begin{equation}
v'(x) = (v \sin i) \, x, \quad -1 \le x \le 1,
\end{equation}
in normalised Cartesian coordinates on the projected stellar disc. Each value of $v'$ therefore corresponds to a chord across the stellar disc perpendicular to the projected equator. The unnormalised kernel is obtained by integrating the surface intensity along these chords,
\begin{equation}
\label{eqn:def_kernel}
K(x) = \int_{-a}^{a} I(x,y) \, dy, \quad a = \sqrt{1-x^2}.
\end{equation}

We assume radial limb-darkening, so that $I(x,y) = I(\mu)$ with $\mu = \sqrt{1 - x^2 - y^2}$. Most commonly used limb-darkening laws can be written as linear combinations of powers of $\mu$:
\begin{equation}
\label{eqn:spec_int_generic}
I(\mu) = \sum_{\gamma} c_{\gamma} \mu^{\gamma}.
\end{equation}
Replacing the above intensity profile into Equation \ref{eqn:def_kernel}, the kernel becomes:
\begin{equation}
\label{eqn:kernel_int_mu}
K(x) = \sum_{\gamma} c_{\gamma} \int_{-a}^{a}  \left ( a^2 - y^2 \right )^{\frac{\gamma}{2}} \, dy, \quad a = \sqrt{1-x^2}.
\end{equation}
Applying the substitution $y= a \cos{t}$:
\begin{equation}
K(x) = \sum_{\gamma} c_{\gamma} a^{\gamma+1} \int_{0}^{\pi}  \left ( \sin{t} \right )^{\gamma+1} \, dt, \quad a = \sqrt{1-x^2}.
\end{equation}
In absence of limb-darkening, there is only one term with $\gamma=0$ ($c_0=1$), so that:
\begin{equation}
K_{\mathrm{unif}}(x) = 2 \sqrt{1-x^2}.
\end{equation}

The linear limb-darkening law coincides with Equation \ref{eqn:spec_int_generic} for $c_0=1-u$ and $c_1=u$. The corresponding unnormalised kernel is:
\begin{equation}
\label{eqn:kernel-linear}
K_{\mathrm{lin}}(x) = 2 (1-u) \sqrt{1-x^2} + \frac{\pi}{2} u (1-x^2).
\end{equation}

The kernel for the quadratic law is obtained with $c_0=1-u_1-u_2$, $c_1=u_1+2u_2$ and $c_2=-u_2$:
\begin{equation}
K_{\mathrm{quad}}(x) = 2 (1-u_1-u_2) \sqrt{1-x^2} + \frac{\pi}{2} (u_1+2u_2) (1-x^2) - \frac{4}{3} u_2 (1-x^2)^{\frac{3}{2}}.
\end{equation}

The integrals for non-integer $\gamma$ exponents do not admit a closed-form analytical solution. They can, however, be expressed in terms of the $\Gamma$ function:
\begin{equation}
\label{eqn:sin_integ_gamma}
\int_{0}^{\pi}  \left ( \sin{t} \right )^{\gamma+1} \, dt = \pi \frac{ \Gamma{ \left ( \frac{\gamma+2}{2} \right ) } }{ \Gamma{ \left ( \frac{\gamma+3}{2} \right ) } }, \quad \gamma > -2.
\end{equation}

By combining Equations \ref{eqn:kernel_int_mu} and \ref{eqn:sin_integ_gamma}, the broadening kernel can be computed for other limb-darkening laws, including the square-root, power-2, and four-coefficient prescriptions.

\section{Numerical aspects of the rotation kernel} \label{appendix-aspects}

\begin{figure}[]
\centering
\includegraphics[width=0.99\hsize]{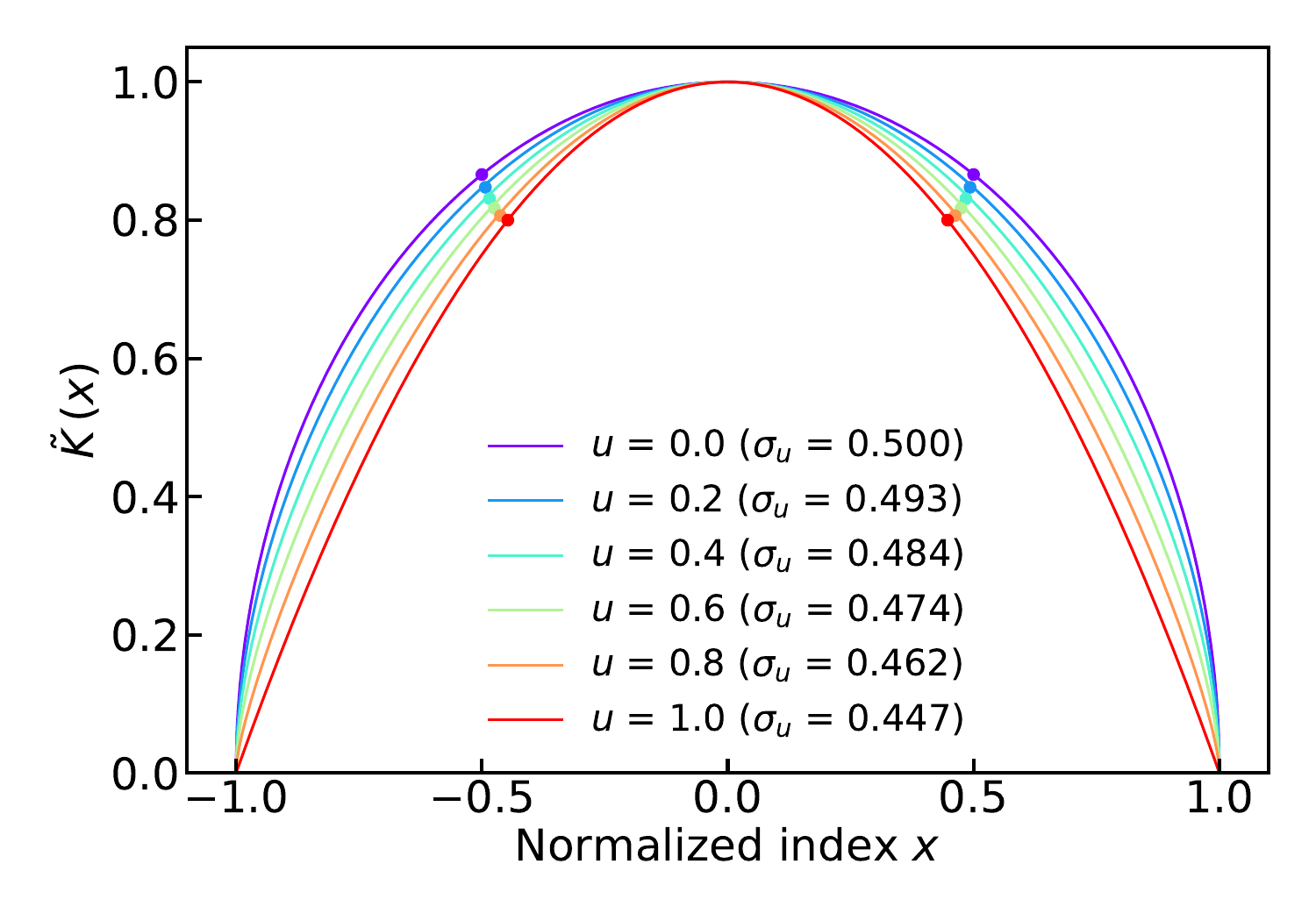}
\caption{ Rotational broadening kernel for different limb-darkening coefficients $u$. Circular markers correspond to the effective width of each kernel.}
\label{fig:kernel_jk}
\end{figure}

Fig.\,\ref{fig:kernel_jk} shows the resulting normalised kernel in Equation \ref{eqn:discrete_kernel} for selected values of the limb-darkening coefficient \(u\). The kernel transitions from a semi-circular profile for \(u = 0\) to an increasingly parabolic shape as \(u \to 1\).
We derived the effective width of each kernel, $\sigma_u$, where $\sigma_u^2$ corresponds to the kernel variance including limb-darkening. The ratio $\sigma_u / \sigma_0$ is a measure of the underestimation of $v \sin i$ when limb-darkening is neglected. For instance, $u = 0.6$ leads to an underestimation of approximately 6\,\%. As illustrated by the narrower kernel at higher $u$, neglecting limb-darkening biases the recovered $v \sin i$ toward lower values.

\begin{figure}[]
\includegraphics[width=0.99\hsize]{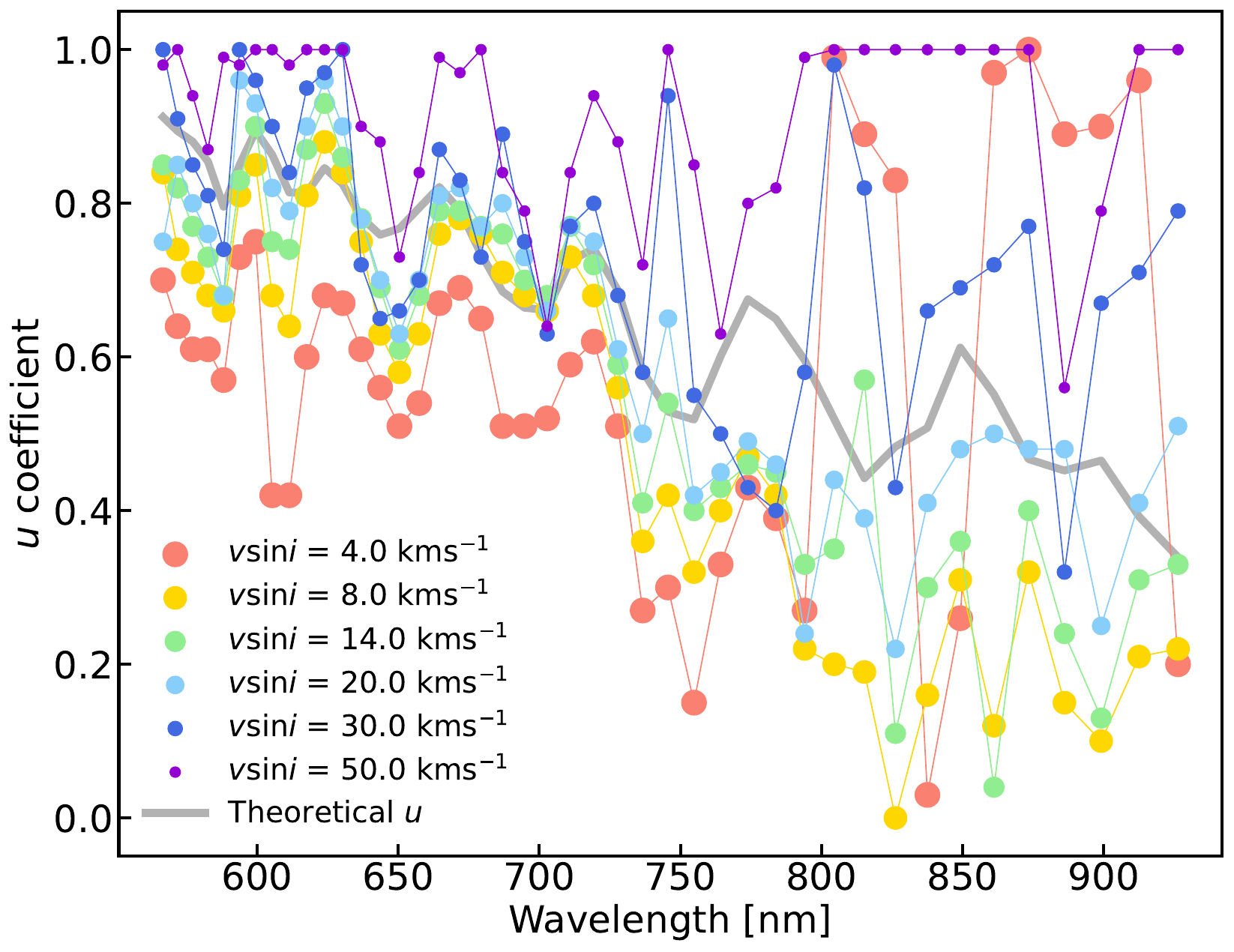}
\caption{Limb-darkening $u$ coefficient as a function of wavelength, when fitting simultaneously with $v \sin i$. 
\label{fig:appendix-u}}
\end{figure}

We attempted to simultaneously fit $v \sin i$ and the limb-darkening coefficient $u$ using the oversampled convolution method, but found notable inconsistencies. The fitted $u$ values vary systematically with the target’s $v \sin i$ (see Fig.\,\ref{fig:appendix-u}), and for higher rotational velocities and $T_{\mathrm{eff}}$, where the method becomes less reliable (Sect.\,\ref{sect:subsection_ni}), $u$ tends to converge toward 1 across all spectral orders, attempting to compensate for discrepancies between the input and recovered $v \sin i$. Furthermore, the fitted coefficients show poor agreement with the theoretical predictions.

\begin{figure}[]
\includegraphics[width=0.99\hsize]{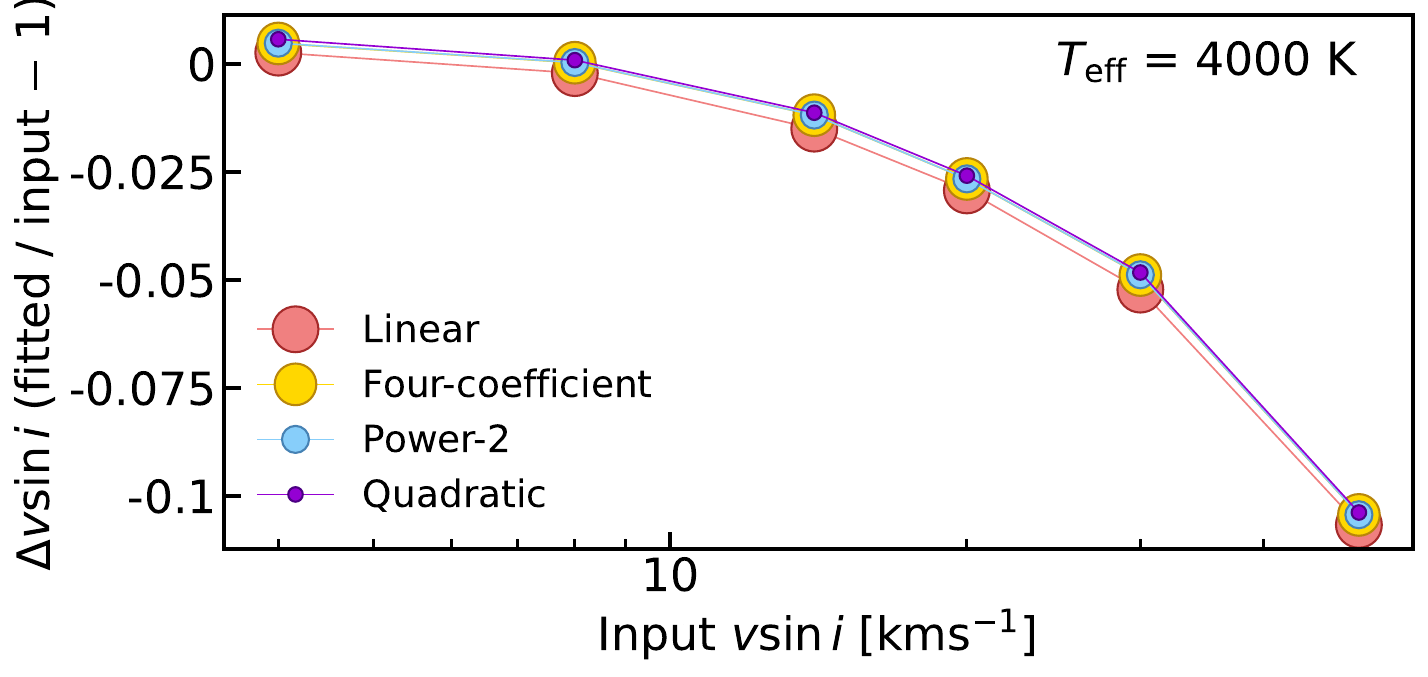}
\caption{Normalized difference between input and fitted $v \sin i$ as a function of the input $v \sin i$ for different limb-darkening laws.
\label{fig:appendix-LD-models}}
\end{figure}

We examined whether the limitations of the method could arise from an overly simplified treatment of limb-darkening. In the context of exoplanetary transit studies, the most commonly employed laws are linear \citep{milne1921}, quadratic \citep{kopal1950}, square-root \citep{diaz1992}, four-coefficient \citep{claret2000}, and power-2 \citep{hestroffer1997}.
Fig.\,\ref{fig:appendix-LD-models} shows the differences between the fitted $v \sin i$ values obtained as a function of the target’s $v \sin i$, for 4000\,K. The discrepancies among the non-linear models are minimal, remaining below 0.5\,$\%$ even in the most extreme cases when compared to the linear law.

\section{Oversampling and limb-darkening impact during convolution}
\label{sec:ld-oversampling}

\begin{table*}[]
\centering
\caption{Derived $v\sin i$ values for all the $T_\mathrm{eff}$ and target $v\sin i$ combinations, using regular convolution without limb darkening (RCWLD) and oversampled convolution with linear limb darkening (OCLLD).}
\label{tab:ld-oversampling}
\begin{tabular}{cccccc}
\hline
\hline
\begin{tabular}[c]{@{}c@{}}Target $v \sin i$ \\{[}\,km\,s$^{-1}${]}\end{tabular} & \begin{tabular}[c]{@{}c@{}}Method\end{tabular} &  \begin{tabular}[c]{@{}c@{}}$v \sin i$ {[}\,km\,s$^{-1}${]}\\ ($T_{\mathrm{eff}}$\,=\,2500\,K)\end{tabular} & \begin{tabular}[c]{@{}c@{}}$v \sin i$ {[}\,km\,s$^{-1}${]}\\ ($T_{\mathrm{eff}}$\,=\,3000\,K)\end{tabular} & \begin{tabular}[c]{@{}c@{}}$v \sin i$ {[}\,km\,s$^{-1}${]}\\ ($T_{\mathrm{eff}}$\,=\,3500\,K)\end{tabular} & \begin{tabular}[c]{@{}c@{}}$v \sin i$ {[}\,km\,s$^{-1}${]}\\ ($T_{\mathrm{eff}}$\,=\,4000\,K)\end{tabular} \\ \hline
\rowcolor[HTML]{EFEFEF} 
4.0 & \begin{tabular}[c]{@{}c@{}} RCWLD \\ OCLLD \end{tabular} & \begin{tabular}[c]{@{}c@{}} 3.97\,$\pm$\,0.31 \\ 3.96\,$\pm$\,0.06  \end{tabular} & \begin{tabular}[c]{@{}c@{}} 4.08\,$\pm$\,0.25 \\ 4.1\,$\pm$\,0.06  \end{tabular} & \begin{tabular}[c]{@{}c@{}} 4.18\,$\pm$\,0.22 \\ 4.09\,$\pm$\,0.07 \end{tabular} & \begin{tabular}[c]{@{}c@{}} 4.26\,$\pm$\,0.22 \\  4.01\,$\pm$\,0.06 \end{tabular} \\
8.0 & \begin{tabular}[c]{@{}c@{}} RCWLD \\ OCLLD \end{tabular} & \begin{tabular}[c]{@{}c@{}} 7.3\,$\pm$\,0.18 \\ 7.86\,$\pm$\,0.14 \end{tabular} & \begin{tabular}[c]{@{}c@{}} 7.77\,$\pm$\,0.27 \\ 8.15\,$\pm$\,0.12 \end{tabular} & \begin{tabular}[c]{@{}c@{}} 7.74\,$\pm$\,0.21 \\ 8.09\,$\pm$\,0.13 \end{tabular} & \begin{tabular}[c]{@{}c@{}} 7.69\,$\pm$\,0.17 \\ 7.98\,$\pm$\,0.15 \end{tabular} \\
\rowcolor[HTML]{EFEFEF}
14.0 & \begin{tabular}[c]{@{}c@{}} RCWLD \\ OCLLD \end{tabular} & \begin{tabular}[c]{@{}c@{}} 12.64\,$\pm$\,0.4 \\ 13.7\,$\pm$\,0.28 \end{tabular} & \begin{tabular}[c]{@{}c@{}} 13.2\,$\pm$\,0.41 \\ 14.15\,$\pm$\,0.18 \end{tabular} & \begin{tabular}[c]{@{}c@{}} 13.35\,$\pm$\,0.3 \\ 13.97\,$\pm$\,0.28 \end{tabular} & \begin{tabular}[c]{@{}c@{}} 13.12\,$\pm$\,0.33 \\ 13.79\,$\pm$\,0.4 \end{tabular} \\ 
20.0 & \begin{tabular}[c]{@{}c@{}} RCWLD \\ OCLLD \end{tabular} & \begin{tabular}[c]{@{}c@{}} 17.95\,$\pm$\,0.59 \\ 19.49\,$\pm$\,0.43 \end{tabular} & \begin{tabular}[c]{@{}c@{}} 18.79\,$\pm$\,0.56 \\ 20.1\,$\pm$\,0.24 \end{tabular} & \begin{tabular}[c]{@{}c@{}} 18.72\,$\pm$\,0.5 \\ 19.73\,$\pm$\,0.59 \end{tabular} & \begin{tabular}[c]{@{}c@{}} 18.55\,$\pm$\,0.73 \\ 19.41\,$\pm$\,0.84 \end{tabular} \\
\rowcolor[HTML]{EFEFEF} 
30.0 & \begin{tabular}[c]{@{}c@{}} RCWLD \\ OCLLD \end{tabular} & \begin{tabular}[c]{@{}c@{}} 26.91\,$\pm$\,0.84 \\ 29.16\,$\pm$\,0.72 \end{tabular} & \begin{tabular}[c]{@{}c@{}} 28.07\,$\pm$\,0.7 \\ 29.83\,$\pm$\,0.49 \end{tabular} & \begin{tabular}[c]{@{}c@{}} 27.57\,$\pm$\,1.36 \\ 29.17\,$\pm$\,1.6 \end{tabular} & \begin{tabular}[c]{@{}c@{}} 26.99\,$\pm$\,1.82 \\ 28.43\,$\pm$\,2.03 \end{tabular} \\
50 & \begin{tabular}[c]{@{}c@{}} RCWLD \\ OCLLD \end{tabular} & \begin{tabular}[c]{@{}c@{}} 44.07\,$\pm$\,1.81 \\ 47.89\,$\pm$\,1.7 \end{tabular} & \begin{tabular}[c]{@{}c@{}} 45.5\,$\pm$\,1.67 \\ 48.62\,$\pm$\,1.85 \end{tabular} & \begin{tabular}[c]{@{}c@{}} 43.72\,$\pm$\,4.96 \\ 46.01\,$\pm$\,5.39 \end{tabular} & \begin{tabular}[c]{@{}c@{}} 42.95\,$\pm$\,5.43 \\ 44.66\,$\pm$\,5.91 \end{tabular} \\

\hline
\end{tabular}
\end{table*}

Table\,\ref{tab:ld-oversampling} presents the results obtained using regular convolution without limb darkening (RCWLD) and oversampled convolution with linear limb darkening (OCLLD) for all combinations of $v\sin i$ and $T_{\mathrm{eff}}$ in our spectral sample. In general, the more sophisticated approach yields values closer to the target $v\sin i$ and with smaller uncertainties. The simplest method systematically underestimates the rotational broadening of the target spectrum and, in most cases, does not agree with the target value within the errors.

\section{Comparison with other works} \label{appendix-comparison}

\begin{figure}
    \centering
    \includegraphics[width=0.9\hsize]{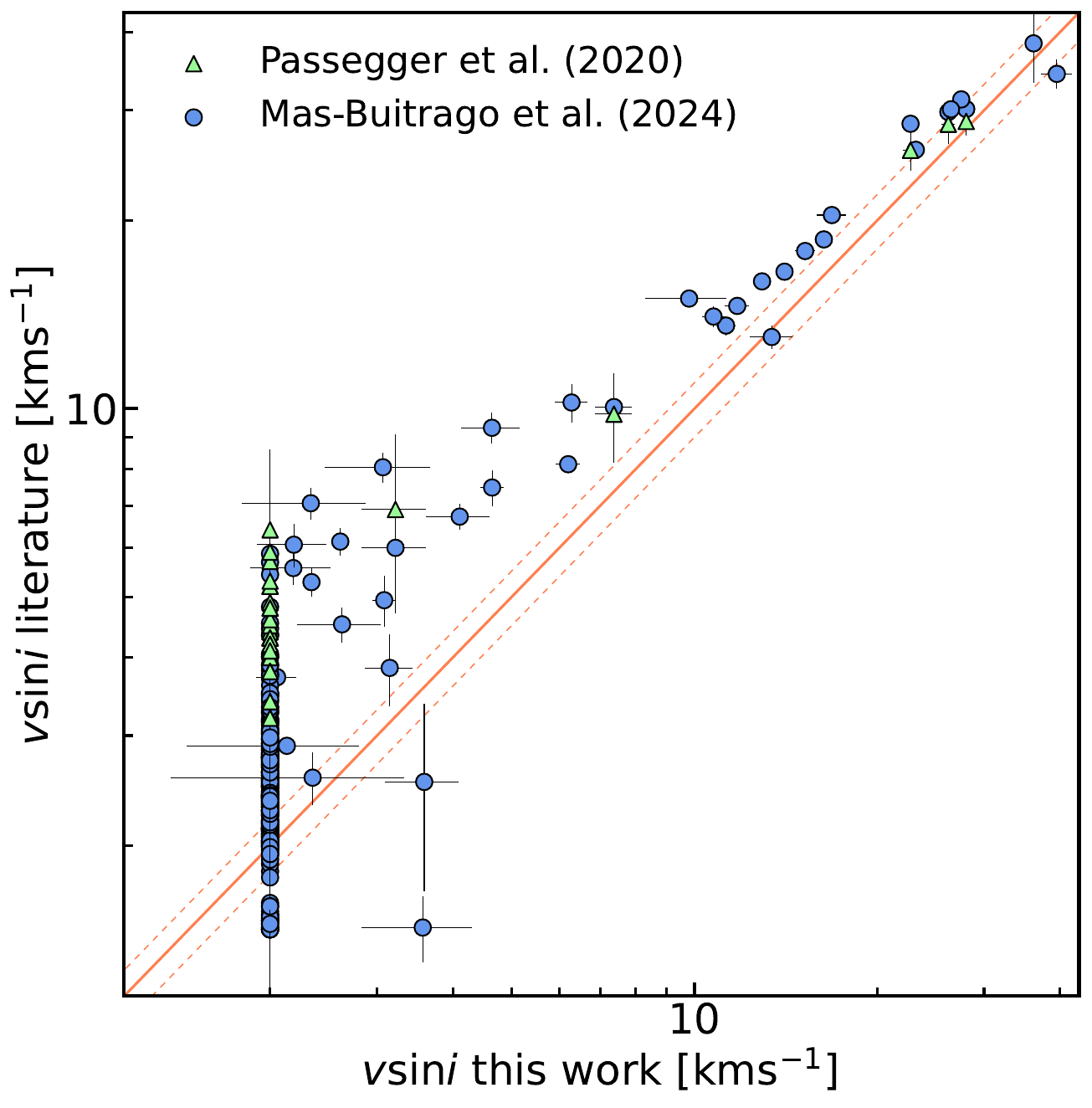}
    \caption{Comparison of our $v\sin i$ values with the results from \citet{passegger2020} and \citet{mas-buitrago2024}}
    \label{fig:vsini_mas24}
\end{figure}

We also compared our catalogue with the results of \citet{passegger2020} and \citet{mas-buitrago2024} (see Fig.\,\ref{fig:vsini_mas24}), who derived the projected rotational velocities of CARMENES targets using deep-learning methods. Their $v\sin i$ values are typically larger, particularly for slow rotators. For $v\sin i$ above 10\,km\,s$^{-1}$, the differences decrease. Both studies also report systematically higher $v\sin i$ values than those derived by \citet{reiners2018}. Our results lie between these estimates, but are closer to those of \citet{reiners2018} and \citet{reiners2022}.

\begin{figure}
    \centering
    \includegraphics[width=0.9\hsize]{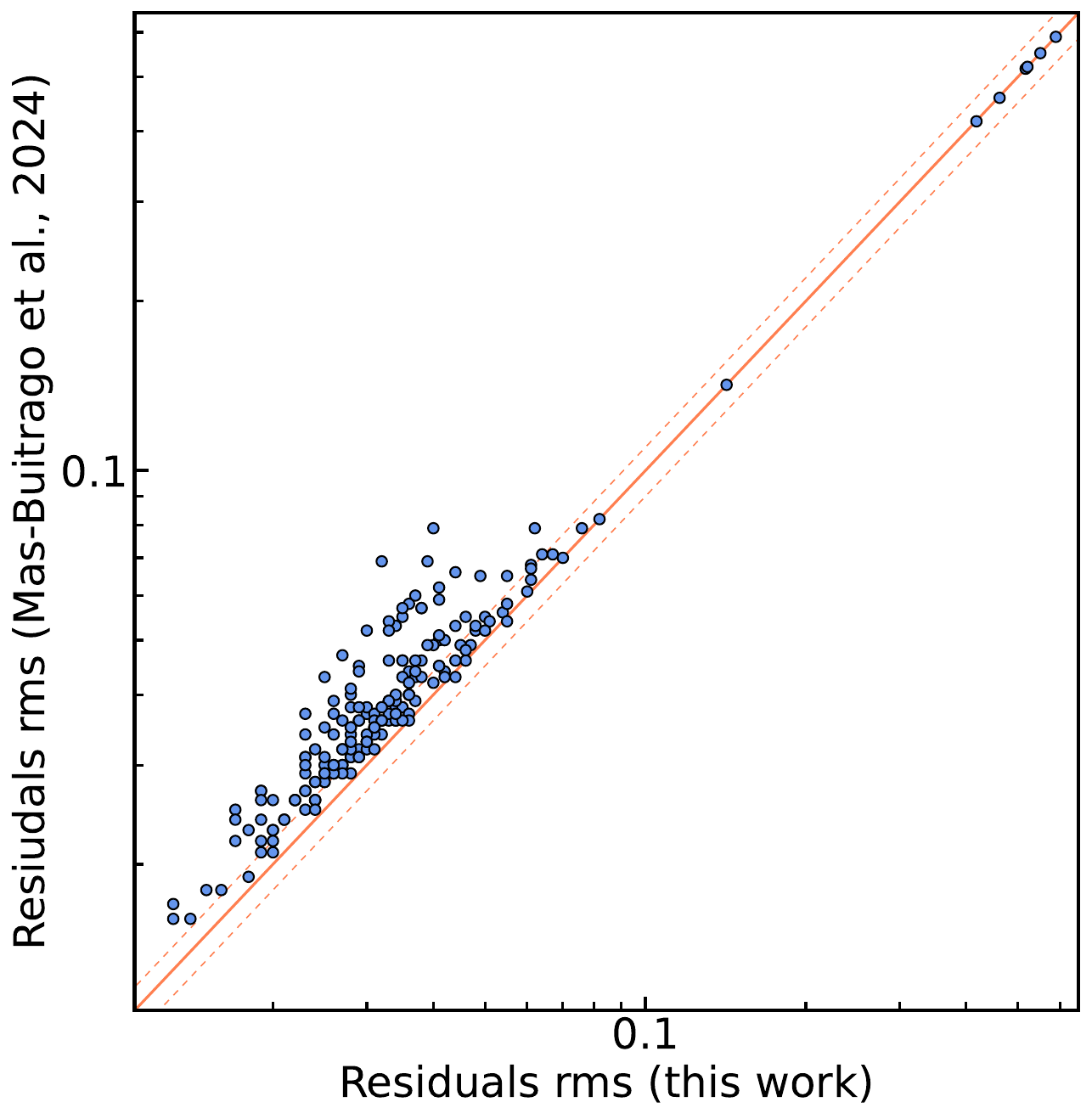}
    \caption{The rms of the residuals between target spectrum and template rotationally broadened using both $v\sin i$ from this work and \citet{mas-buitrago2024}.}
    \label{fig:rms_mas24}
\end{figure}

For all targets in common with \citet{mas-buitrago2024}, we compared the stellar spectrum with the corresponding template (see Table\,\ref{tab:tpls}) rotationally broadened using both sets of $v\sin i$ values. The rms of the residuals is shown in Fig.\,\ref{fig:rms_mas24}. In general, our $v\sin i$ measurements yield smaller rms values, particularly for slow rotators. This indicates that our results reproduce the observed broadening of the spectral lines in the CARMENES VIS spectra more accurately.

\section{Spectrum and $v \sin i$ of J14155+046} \label{appendix-J14155}

\begin{figure}[h]
\includegraphics[width=0.99\hsize]{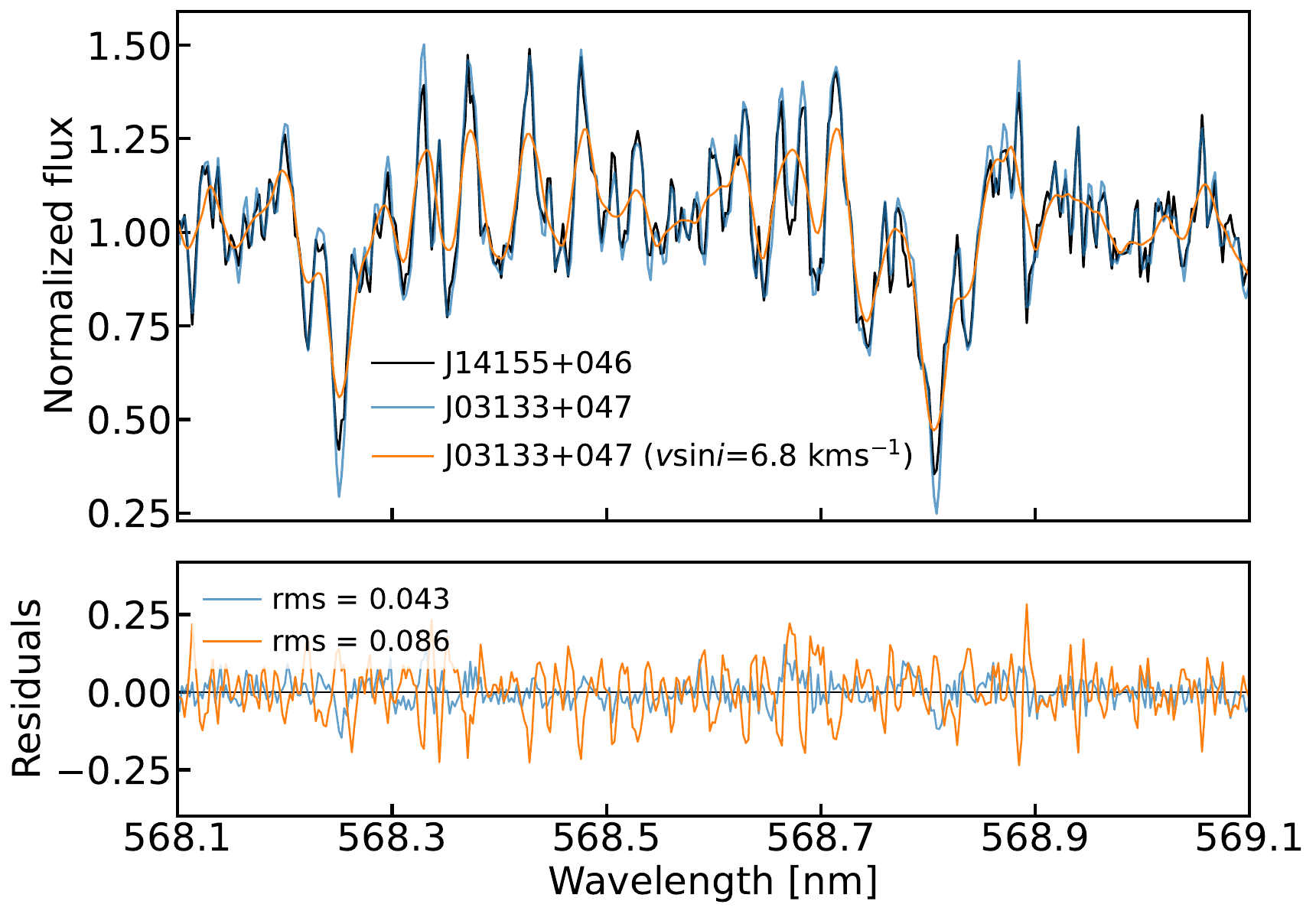}
\caption{Top: chunk of the CARMENES-VIS co-added spectrum of J14155+046 compared to the spectrum of J03133+047, without and with 6.8\,km\,s$^{-1}$ rotational broadening \citep{jenkins2009}.  Bottom: residuals with respect to J14155+046.
\label{fig:spectra_J14144}}
\end{figure}

\citet{jenkins2009} reported a projected rotational velocity of $v \sin i = 6.8$\,km\,s$^{-1}$ for J14155+046. In our analysis, we obtained a value below 2\,km\,s$^{-1}$, which corresponds to the lower limit of our method. Comparing the spectrum of J14155+046 with that of J03133+047, a reference star in our sample with $v \sin i < 2$\,km\,s$^{-1}$, we find a better match than when broadening the spectrum of J03133+047 to higher velocities, as the rms of the residuals halve (see Fig.\,\ref{fig:spectra_J14144}).

\end{appendix}

\end{document}